\documentclass[iop,appendixfloats]{emulateapj}

\newcommand{\figurepath}{.}

\newcommand{\rsun}{{\rm R}_{\sun}}

\newbox\grsign \setbox\grsign=\hbox{$>$}
\newdimen\grdimen \grdimen=\ht\grsign
\newbox\laxbox \newbox\gaxbox
\setbox\gaxbox=\hbox{\raise.5ex\hbox{$>$}\llap
     {\lower.5ex\hbox{$\sim$}}}\ht1=\grdimen\dp1=0pt
\setbox\laxbox=\hbox{\raise.5ex\hbox{$<$}\llap
     {\lower.5ex\hbox{$\sim$}}}\ht2=\grdimen\dp2=0pt

\shorttitle{Jet launching in binaries}
\shortauthors{Sheikhnezami \& Fendt}
\usepackage[usenames,dvipsnames]{color}
\usepackage{amsmath}
\usepackage{float}
\usepackage{ctable} 
\usepackage{graphicx}
\begin{document}

\title{
Long-term simulation of MHD jet launching from an orbiting star-disk system}
\author{Somayeh Sheikhnezami
\altaffilmark{1,2} 
\&
        Christian Fendt
 \altaffilmark{2}
}
\altaffiltext{1}{School of Astronomy, Institute for Research in Fundamental Sciences (IPM), P.O. Box 1956836613, Tehran, Iran}
\altaffiltext{2}{Max Planck Institute for Astronomy, K\"onigstuhl 17, DE-69117 Heidelberg, Germany}
\email{snezami@ipm.ir, fendt@mpia.de}                                   

\begin{abstract}
We present fully three-dimensional magnetohydrodynamic jet launching simulations of a 
jet source orbiting in a binary system.
We consider a time-dependent binary gravitational potential, thus all tidal forces that 
are experienced in the non-inertial frame of the jet-launching primary.
We investigate systems with different binary separation, different mass ratio, and 
different inclination between the disk plane and the orbital plane.
The simulations run over a substantial fraction of the binary orbital period.
All simulations show similar local and global non-axisymmetric effects such as local instabilities in 
the disk and the jet, or global features such as disk spiral arms and warps, or a global re-alignment 
of the inflow-outflow structure.
The disk accretion rate is higher than for axisymmetric simulations, most probably 
due to the enhanced angular momentum transport by spiral waves.
The disk outflow leaves the Roche lobe of the primary and becomes disturbed by tidal effects.
While a disk-orbit inclination of $10\degr$ still allows for a persistent outflow, an inclination 
of $30\degr$ does not, suggesting a critical angle in between.
For moderate inclination we find indication for jet precession such that the jet axis 
starts to follow a circular pattern with an opening cone of $\simeq 8\degr$.

Simulations with different mass ratio indicate a change of time scales for the tidal forces to 
affect the disk-jet system.
A large mass ratio (a massive secondary) leads to stronger spiral arms, a higher (average) accretion 
and a more pronounced jet-counter jet asymmetry.

\end{abstract}
\section{Introduction}
Astrophysical jets emerge from magnetized accretion disk systems.
It is now commonly accepted that magnetohydrodynamic (MHD) processes are essential for the 
launching, acceleration and collimation of the outflows and jets from accretion disks
\citep{2007prpl.conf..277P, 2015SSRv..191..441H}.
The overall idea is that energy and angular momentum are extracted from the disk relying 
on an efficient magnetic torque that is essentially provided by a global, i.e. large-scale 
(jet) magnetic field threading the disk. 
When the inclination of the field lines is sufficiently small ($< 60 \degr$ for a cold wind), 
magneto-centrifugal forces can accelerate the matter along the field line, efficiently {\em forming} 
a high speed outflow \citep{1982MNRAS.199..883B, 1983ApJ...274..677P}.
Magnetohydrodynamic forces are responsible for {\em launching} the outflow, i.e. initiating the 
upward motion of disk material toward the disk surface where it is feeding the outflow
\citep{1997A&A...319..340F}.

A number of MHD simulations have investigated the time-dependent jet launching including the
time-evolution of the resistive accretion disk
\citep{2002ApJ...581..988C, 2007A&A...469..811Z, 2010A&A...512A..82M, 2012ApJ...757...65S, 
2014ApJ...796...29S, 2016ApJ...825...14S}.
However, yet, it is not fully understood which kind of disks do launch jets and over what time scales 
such a mechanism works.
Recent three-dimensional (3D) ideal MHD simulations of jet launching consider in particular the 
interplay between the large-scale magnetic field outside the disk and the tangled field structure inside 
the disk \citep{2017arXiv170104627Z}.
This is a central question for jet launching.

There is evidence that jets are also formed in binary systems as observations indicate
jet precession or a ballistic three-dimensional jet motion
\citep{2000ApJ...535..833S, 2002MNRAS.335.1100C, 2004HEAD....8.2903M, 2007A&A...476L..17A, 2016A&A...593A.132P, Beltran2016}.
We note that it is well known that young stars often form in binary or even multiple systems.
Examples of numerical simulations 
on ballistic jet motion of jets from binaries are \citet{2009ApJ...707L...6R, 2013MNRAS.428.1587V}. 
Jets are also found being ejected from evolved stars.
One of the rare examples is the asymptotic giant branch star W43A \citep{2006Natur.440...58V}.
Collimated relativistic outflows were found from a number of compact binary systems, so-called 
micro-quasars \citep{1979ApJ...233L..63M, 1999ARA&A..37..409M, 2015A&A...584A.122L}.

The structure and evolution of disks in binary systems has been studied for long time.
In interacting binary systems the accretion disk around the primary feels the tidal torques
exerted by the secondary.
Seminal papers have investigated the gravitational interaction between the circum-stellar disks and the binary
stars
and in particular have derived limits for an outer disk radius until which the disk is in a quasi-equilibrium
(see e.g. \citealt{1977MNRAS.181..441P, 1977ApJ...216..822P, 1994ApJ...421..651A}).
The latter authors compare analytical estimates of the gravitational interaction between the disks and the binaries
applying smoothed particle hydrodynamics simulations.
In general, these works find disk radii of typically 0.4-0.5 times the semi major axis, depending on the system 
parameters, such as mass ratio or eccentricity.

A more recent work following this approach is \citet{2007MNRAS.376...89T} who confirm disk truncation by viscous 
angular momentum transport in close binary systems by performing hydrodynamical simulations of viscous
accretion disks for different binary mass ratio.
In fact, the Roche lobe is expected to be the maximum disk radius while material outside this radius will be lost from 
the system.
In \citet{2008A&A...487..671K} eccentric disks in close binary systems were studied by performing 2D
hydrodynamic viscous simulations, finding that the disk aspect ratio as well as the mass transfer rate may have 
substantial impact on the formation of an eccentric disk and disk precession.
Applying 3D hydrodynamical simulations to study the complex disk structure arising in misaligned binaries 
\citet{2010A&A...511A..77F} investigate the specific conditions that lead to inclined disks.
They find that disks that are thinner but have a higher viscosity
can develop a significant twist before achieving a rigid-body precession.
For very thin disks, these disks may brake up or can be disrupted by a strong differential precession.

The first 3D magnetohydrodynamic simulations of a circum-binary disk surrounding an equal-mass system were 
performed by \citet{2012ApJ...749..118S}.
A recent paper studying the disk evolution in close binaries is \citet{2017ApJ...841...29J}.
These authors perform global 3D MHD simulations studying the relative importance of spiral shocks and the 
magnetorotational instability (MRI) for angular momentum transport - in particular their dependence on the geometry 
and strength of the seed magnetic field and the Mach number of the disk. 

To study the dynamics and time evolution of {\em jets launched in binary systems} is the major aim of this
paper. We are concerned about the following questions.
How does the alignment of the jet changes in time if the orbital plane of the jet launching accretion disk
and that binary orbital plane are not co-planar?
Is there an upper limit for the disk-orbital plane inclination beyond which 3D effects prevent
persistent jet formation?
Is there indication for tidal effects such as jet precession?
Naturally, our study will be limited by numerical and physical constraints if compared to e.g. 2D jet launching simulations
or hydrodynamic binary disk simulations that are both numerically less expensive.

In the present paper we extend our previous study \citep{2015ApJ...814..113S} such that we now perform the
simulations for  
(i) a longer integration time, up to more than half of a binary orbit, and apply
(ii) a time-dependent 3D gravitational potential that is acting on the disk and the jet that is ejected 
from the disk.

Our paper is structured as follows.
Section 2 specifies some of observations of jets formed in binary systems.
Section 3 describes our model setup i.e. initial conditions and  boundary conditions for the binary star- disk-jet system.
In Section 4 we present the results of the long-term evolution of jet launching along the orbit of a binary system.
In Section 5 we compare some observational numbers for precessing jets from binary systems in respect to our model setup.
Section 6 summarizes the results.
In the appendix we discuss the Blandford-Payne criterion in respect to a 3D gravitational potential.

\section{Observational evidence of precessing jets}
\label{sec:observations}
Observations have detected a number of sources with jets deviating from a straight direction of propagation
that potentially can be interpreted as due to precession.
A possible explanation for such features is that these jet emerge in a binary or even multiple system.

Among the confirmed binary systems that are sources of jets are T\,Tau 
\citep{1997A&AS..126..437H, 2002ApJ...568..771D, 2003AJ....125..858J} 
or RW\,Aur \citep{1996AJ....111.2403H, 2012ARep...56..686B}.
Another example is the spectroscopically identified bipolar jet of the pre-main sequence binary 
KH\,15D  that seems to be launched from the innermost part of the circum-binary disk, or may, 
alternatively, result from the merging of two outflows each of them driven by the individual
stars, respectively \citep{2010ApJ...708L...5M}.
The existence of a circum-binary disk in KH\,15D is evident from dust settling \citep{2010ApJ...711.1297L}.
The  disk in KH\,15D  is tilted, warped, and seems to be precessing with respect to the binary orbit
\citep{2004AJ....127.2344J, 2004ApJ...607..913C, 2004AJ....128.1265J, 2004ApJ...603L..45W}.

\cite{Beltran2016} studied another source with jet precession which is the well-known 
high-mass young stellar object G35.20−0.74N. 
Their VLA observations have revealed the presence of a binary system located at the 
geometrical center of the radio jet. 
This binary system, associated with a Keplerian disk, consists of two B-type stars of 11 
and 6\,$M \odot$. 
The authors argue that the precession induced in the binary system is the main reason of 
the S-shaped morphology of the radio jet observed in this object.  
The effect of an intrinsic binary motion on the large scale jet geometry has been investigated by 
\citet{Fendt1998}, discussing a S-shaped or C-shaped jet geometry for orbiting jet sources.

The high-mass protostar NGC\,7538\,IRS1 is another outflow source.
\citet{2006A&A...455..521K} studied the possibility of outflow precession and show that the triggering mechanism 
might be the non-co-planar tidal interaction of a close companion with the circum-binary protostellar disk.
Their observations resolve this nearby massive protostar as a close binary with a separation of 195 mas.

Another example of a binary star is HK\,Tau.
Here, both stars, HK\,Tau\,B and HK\,Tau\,A, have a circum-stellar disk.
Both disks are misaligned with respect to the orbital plane of the binary \citep{2014Natur.511..567J}. 
There is not yet a clear observation of jets in this source, however, it is a system that potentially 
may show jet precession.

An exceptionally striking example is the well-known X-ray binary SS\,433
\citep{1979ApJ...233L..63M, 2015A&A...574A.143M}, which has a relativistic jet that is precessing.
VLBA observations of SS433 at $10^{-4}$ pc scale covering 40 days of data show the dynamics and the precession 
of the jet close to its launching area \citep{2004HEAD....8.2903M}.
\citet{2015A&A...574A.143M} studied the SS433 structure at different scales by 3D hydrodynamic simulations 
and subsequent radiation transfer to investigate the discrepancy between the larger scales of the jet of 
SS433 and its inner region. 
 
\begin{figure}[t]
\centering
\includegraphics[width=1.\columnwidth]{\figurepath/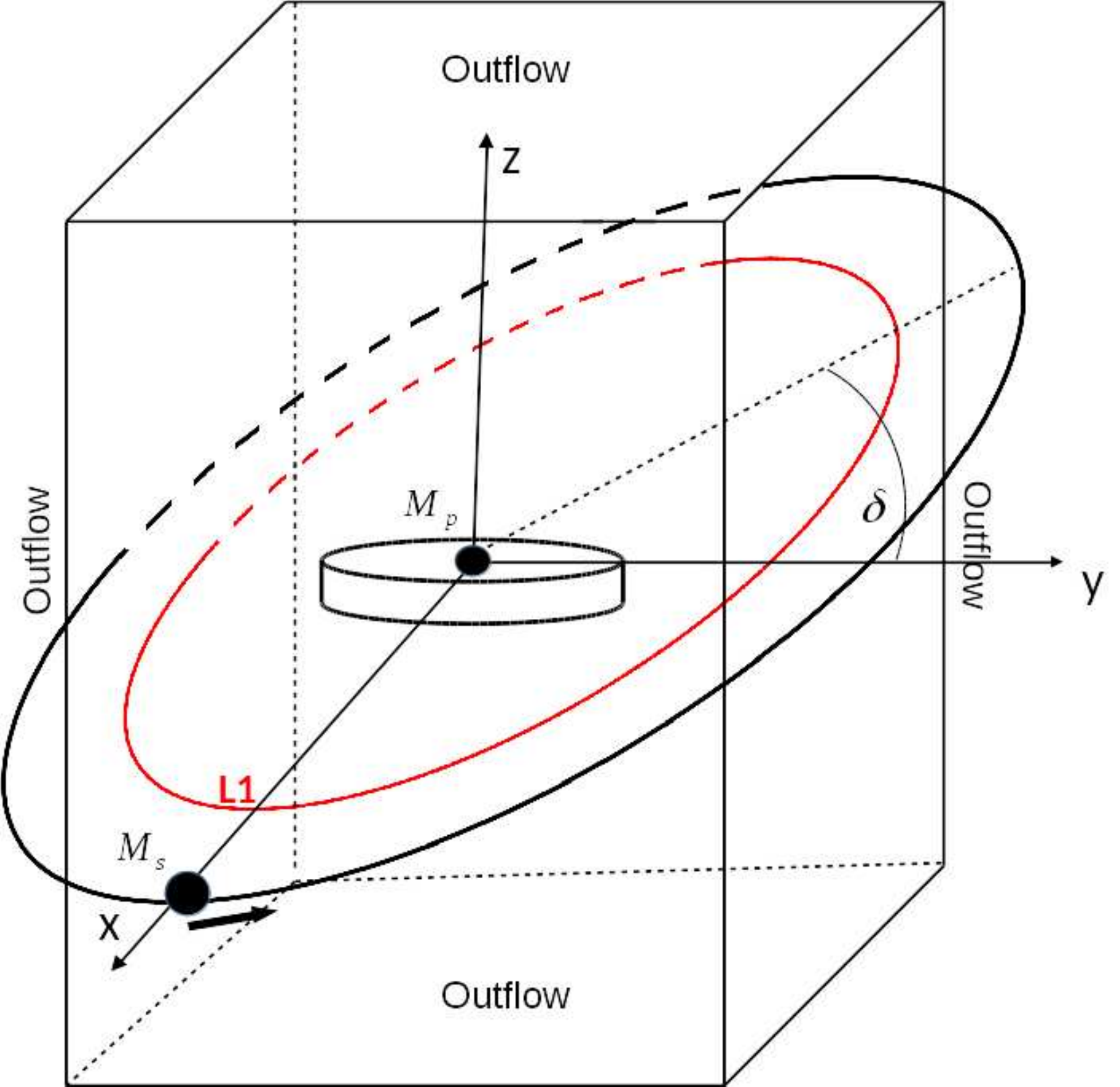}
\caption{Sketch of our model setup. 
We consider a binary system consisting of a primary of mass $M_{\rm p}$ surrounded by an accretion 
disk of size $r_{\rm out}$ and a secondary of mass $M_{\rm s}$. 
The Cartesian coordinate system originates in the primary.
The orbital plane is inclined with an angle $\delta$ with respect to the initial accretion disk mid-plane.
The secondary is located outside the computational domain, separated by a distance $D$ from the primary. 
The red circle displays the orbit of the inner Lagrange point L1 with respect to the primary.
For a mass ratio  $q \equiv M_{\rm s} / M_{\rm p}$ of unity, the L1 is located halfway between 
the two stars.
 }
\label{fig:binary_cartoon2}
\end{figure}

\section{Model approach for jets from binary systems}
We consider a binary system with a primary of mass $M_{\rm p}$ and a secondary of mass $M_{\rm s}$,
separated by the distance $D$.
The primary is surrounded by a disk of initial size $r_{\rm out}$.
The location of the secondary is chosen to be outside the computational domain.
The orbital plane of the binary system is inclined towards the initial accretion disk
by an angle $\delta$.
The Lagrange points L1, L2 and L3 are outside the initial disk radius.
The Lagrange points L1 and L3 could be located in the computational domain,
 and even within the initial accretion disk depending on mass ratio and separation.
 
Figure \ref{fig:binary_cartoon2} illustrates the general setup for our simulations of 
MHD jet launching from a circumstellar disk located in a binary system.
The computational domain is shown by a rectangular box where we have indicated the inner Lagrange point L1 
of the binary and the orbit of the secondary. 
The L1 point orbits with the secondary (red circle).
In the present paper, we implement the full gravitational potential of the binary in
the initially axisymmetric setup of the disk-jet.
In other words, we take into account the orbital motion of the binary.

Since we choose the primary as the origin of the computational domain, we do not consider an inertial frame.
The advance of this work compared to our previous paper \citep{2015ApJ...814..113S} is that
all tidal terms induced by the non-inertial frame are considered.

\subsection{Magnetohydrodynamic equations}
We apply the MHD code PLUTO \citep{2007ApJS..170..228M, 2012ApJS..198....7M}, version 4.0, to 
solve the time-dependent, resistive, in-viscous MHD equations, namely for the
conservation of mass, momentum, and energy,
\begin{equation}
\frac{\partial\rho}{\partial t} + \nabla \cdot(\rho \vec v)=0,
\end{equation}
\begin{equation}
\frac{\partial(\rho \vec v)}{\partial t} + 
\nabla \cdot \left(\vec v \rho \vec v - \frac{\vec B \vec B}{4\pi} \right) + \nabla \left( P + \frac{B^2}{8\pi} \right)
+ \rho \nabla \Phi = 0,
\end{equation}
Here, $\rho$ is the mass density, $\vec v$ is the velocity, $P$ is the thermal gas pressure,
$\vec B$ stands for the magnetic field,
and $\Phi$ denotes the gravitational potential.

The major advance of this paper is that we consider a time-dependent, three-dimensional gravitational
potential $\Phi(x,y,z,t)$ that represents the Roche potential of a binary system 
(see Sect.~\ref{sec:gravity}). 

The electric current density $\vec j$ is given by Amp\'ere's law $\vec j = (\nabla \times \vec B) / 4\pi$.
The magnetic diffusivity is defined most generally as a tensor $\bar{\bar{\eta}}$.
In this paper, for simplicity we assume a scalar, isotropic magnetic diffusivity $\eta_{ij} \equiv \eta(x,y,z)$
(see Section 3.2.).
The evolution of the magnetic field is described by the induction equation,
\begin{equation}
\frac{\partial \vec B}{\partial t} - \nabla\times (\vec v \times \vec B - \eta \vec j) = 0.
\end{equation}

The cooling term $\Lambda$ in the energy equation can be expressed in terms of ohmic heating
$\Lambda = g \Gamma$, with $\Gamma = ({\eta} \vec j) \cdot \vec j$, and with $g$ measuring 
the fraction of the magnetic energy that is radiated away instead of being dissipated locally. 
For simplicity, again we adopt $g=1$, thus we neglect ohmic heating for the dynamical evolution of
the system.
The total energy density is
\begin{equation}
e = \frac{P}{\gamma - 1} + \frac{\rho v^2}{2} + \frac{B^2}{2} + \rho \Phi.
\end{equation}
The gas pressure follows a polytropic equation of state $P = (\gamma - 1) u$ with 
$\gamma = 5/3$ and the internal energy density $u$.

{
\subsection{Magnetic diffusivity}
Considering {\em resistivity} is essential for jet launching simulations
\citep{2002ApJ...581..988C,2007A&A...469..811Z,2010A&A...512A..82M,2012ApJ...757...65S,
            2013ApJ...774...12F,2014ApJ...793...31S,2016ApJ...825...14S}.
Accretion of disk material across a large-scale magnetic field that is threading 
the disk can only happen if that matter can diffuse across the field.
After some time an equilibrium situation will be established between inward advection 
of magnetic flux along the disk and diffusion of flux in outward direction 
(see e.g. \citealt{2012ApJ...757...65S}).
Essentially, the launching of an outflow is a consequence from re-distributing matter 
across the magnetic field,
and therefore depends strongly on the level of magnetic diffusivity.

In previous works, we have investigated in detail how the dynamics of the
accretion-ejection structure - for example the corresponding mass fluxes, the jet 
rotation, or the jet propagation speed - depends on the profile and the magnitude of
magnetic diffusivity \citep{2012ApJ...757...65S, 2013ApJ...774...12F}.
These papers have typically applied a magnetic diffusivity $\eta = \eta(r,z) \propto h(r,z)$ that
is constant in time and follows a Gaussian vertical profile $h(r,z)$, essentially
parameterized by the the disk thermal scale height $H$.

When we extended our approach to 3D simulations \citep{2015ApJ...814..113S},
we found, however, that such a diffusivity profile may lead to instabilities in the 3D evolution 
of the system.
The most stable and smooth evolution of the accretion-ejection structure we observed when
applying a {\em constant background diffusivity} (as e.g. applied by \citealt{2003A&A...398..825V}).
Therefore, we have followed this approach also for the present paper, and prescribe a constant
level for the magnetic diffusivity inside the disk and for the nearby disk corona,
\begin{equation}
  h(z) =  
\begin{cases}
     \eta_0,  & {\rm for} \quad |z| < 10, \\
      0,      & {\rm for} \quad |z| > 10,
\end{cases}
\label{eta_profile}
\end{equation}
while we assume ideal MHD for the rest of the grid.
We choose $\eta_0 = 0.03$ for all simulations presented here.
We note that the very mass flux rates and other dynamical variables may depend on the exact
diffusivity profile, however, a comparison between simulations applying the same diffusivity profile
can of course be made.
}

\begin{table} 
\caption{Characteristic parameters of our simulation runs. 
Here, $\beta_{\rm i}$ is the initial (maximum) plasma-beta at the inner disk radius,
$D$ is the binary separation (in the orbital plane), 
$\delta$ is the inclination angle between the binary orbit and the initial accretion disk mid-plane,
$q\equiv M_s/M_p$ is the mass ratio between secondary and primary,
$L1$ and $L3$ denote the radial location of the Lagrange points (in the orbital plane), 
$T_{\rm b}$ indicates the orbital period (in time units $t_{\rm i}$),
and $r_{\rm out}=65$ is the initial outer radius of the disk.
All values are given in code units.}
\begin{center}
\begin{tabular}{cccccccr}
\hline
\hline
\noalign{\smallskip}
Run  & $\beta_{\rm i}$ & D & $\delta$ & $ q $ & $L1$ &  $L3$ & $T_{\rm b}$\\
\noalign{\smallskip}
\hline
\noalign{\smallskip}
\noalign{\smallskip}
{\em d150a0}        & 20    & 150    &  0   & 1      &  75      &  -105     &{ 8162} \\
{\em d150a30}       & 20    & 150    & 30   & 1      &  75      &  -105     &{ 8162} \\
{\em d150a10}       & 20    & 150    & 10   & 1      &  75      &  -105     &{ 8162}  \\
{ \em d150m0.5}     & 20    & 150    & 10   & 0.5    & 86       &  -120     &{ 9424}\\
{ \em d150m2}       & 20    & 150    & 10   & 2      & 64       &  -87      & { 6664}\\
{\em d200a30}       & 20    & 200    & 30   & 1      &  100     &  -140     & { 12566}\\
{\em hydro}         & $10^7$ & 150   & 10   & 1      &  75      &  -105     & { 8162}\\
  \noalign{\smallskip}
 \hline
 \noalign{\smallskip}
 \multicolumn{8}{l}{\textsuperscript{*}\footnotesize{The {\em hydro} simulation applies a very weak magnetic field}}
 \end{tabular}
 \end{center}
\label{tbl:1}
\end{table}

\subsection{Numerical setup}
\label{numerical_setup}
Our simulations are performed in 3D Cartesian coordinates $(x,y,z)$.
Note that contrary to our previous axisymmetric simulations, the $z$-axis is not a symmetry axis anymore.
The computational domain typically extends over $ x\in [-80,80]$ and $y\in [-80,80]$
in units of the inner disk radius $r_{\rm i}$ but has different extent in vertical direction $z$.

Cartesian coordinates may cause problems when treating rotating objects (see \citealt{2015ApJ...814..113S}).
Spherical coordinates are well suited for 3D disk simulations
(see e.g. \citealt{2011ApJ...735..122F,2014ApJ...784..121S}), in particular if the region along the vertical
axis is not considered.
However, when investigating the 3D structure of a jet, artificial symmetry constraints by the rotational 
axis boundary conditions must be avoided.

The origin of the coordinate system is located in the primary.
The $z$ axis is chosen to be along the rotation axis of the unperturbed disk.
The accretion disk mid-plane initially follows the $x-y$ plane for $z=0$.
We prescribe the orbital motion of the binary in a plane which has the inclination angle $\delta$ 
with respect to the initial accretion disk mid-plane.
We denote the orbital angular velocity of the secondary around the primary with $\omega$, 
equivalent to the orbital angular velocity of the binary considering the center of mass.

We apply a uniform grid of $200 \times 200 \times 200 = 8\times 10^6$ cells for the very inner part of 
the domain, $-5.0 < x,y,z < 5.0$, in order to optimize the numerical resolution in the jet launching area.
This corresponds to a resolution of about 20 cells per disk scale height ($\Delta x =\delta y= \delta z = 0.05$).
For the rest of the domain, i.e. the range $|\pm 5| < x,y,z<|\pm 80|$, we apply a stretched grid 
of $476\times 476\times 476$ grid cells.
We apply a stretching factor of about 1.01.
We also require that the shape of the grid cells is only moderately stretched,
avoiding ill-defined cell aspect ratios that will limit the convergence of the numerical scheme.

The same normalization as in \citet{2015ApJ...814..113S} is applied.
Distances are expressed in units of the inner disk radius $r_{\rm i}$,
while $P_{\rm d,i}$ and $\rho_{\rm d,i}$ are the disk pressure and density at this radius, 
respectively\footnote{The index {'}i{'} refers to the value at the inner disk radius 
at the equatorial plane at time $t=0$}.
Velocities are normalized in units of the Keplerian velocity $v_{\rm K,i}$ at the inner disk 
radius. 
We adopt $v_{\rm  K,i} = 1$ and $\rho_{\rm d,i} = 1$ in code units.
The pressure is given in units of $P_{\rm d,i} = \epsilon^2 \rho_{\rm d,i} v_{\rm K,i}^2$.
Here, $\epsilon$ is the ratio of the isothermal sound speed to the Keplerian speed, both evaluated at disk 
mid-plane, $\epsilon  \equiv {c_{\rm s}}/{v_{\rm K}}$. 
The magnetic field is measured in units of $B_{\rm i} = B_{z,\rm i}$ where
$B_{\rm i} =\epsilon \sqrt{2/\beta }$ with the plasma-beta $\beta$ as the ratio of the thermal to the magnetic 
pressure\footnote{In PLUTO the magnetic field is normalized considering $4\pi \equiv 1$}.

The dynamical time unit for the simulation is defined by the Keplerian speed at the inner disk radius,
$t_{\rm i} = r_{\rm i} / v_{\rm K,i}$.
Therefore, $t_{\rm i} = T_{\rm K,i}/2\pi$ with the Keplerian period of the disk
at the inner disk radius, $T_{\rm K,i} = 2\pi$ in code units. 

We apply the method of constrained transport (CT) for the magnetic field evolution conserving
$\nabla \cdot B$ by definition.
For the spatial integration we use a linear algorithm with a second-order interpolation scheme,
together with the third-order Runge–Kutta scheme for the time evolution. 
Further, a HLL Riemann solver is chosen.

\subsection{Initial and boundary conditions}
We apply the same initial conditions and boundary conditions as in \citet{2015ApJ...814..113S}.
We prescribe an initially geometrically thin disk with the thermal scale height $H$ and
$\epsilon = H/r =  0.1$. 
The disk is supposed to be in vertical equilibrium between the thermal pressure and the gravity of 
the primary.

The initial disk density distribution is
\begin{equation}
 \rho_{\rm d} = \rho_{\rm d,i}\left(\frac{2}{5\epsilon^2}\left[\frac{r_i}{R}-
          \left(1-\frac{5\epsilon^2}{2}\right) \frac{r_i}{r}\right] \right)^{3/2},
\end{equation}
while for the initial disk pressure distribution we apply
\begin{equation}
P_{\rm d} = P_{\rm d,i} \left(\frac{\rho_{\rm d,i}}{\rho_{\rm d}}\right)^{5/3}.
\end{equation}

Here, $r=\sqrt{x^2+y^2}$ and $R=\sqrt{x^2+y^2+z^2}$ denote the cylindrical and the spherical radius, respectively.
The accretion disk is set into a slightly sub-Keplerian rotation accounting for the radial gas pressure gradient 
and advection and the non-force free structure of the magnetic field, initially.

The initial magnetic field distribution is prescribed by the magnetic flux function $\psi$,
\begin{equation}
\psi(x,y,z)=\frac{3}{4} B_{\rm z,i} r_{\rm i}^2\left(\frac{r}{r_{\rm i}}\right)^{3/4}
\frac{m^{5/4}}{{\left(m^2+\left({z}/{r}\right)^2\right)}^{5/8}},
\end{equation}
where the parameter $m$ determines the magnetic field bending \citep{2007A&A...469..811Z}.
In our model setup $m=0.4$.
Here, $B_{z,\rm i}$ denotes the vertical magnetic field at the inner disk radius, 
$(r=r_{\rm i},z=0)$.
Numerically, the poloidal field components are implemented by prescribing the magnetic vector potential $A_{\phi}(x,y) = \psi/r$.
Initially $B_{\phi}=0$. 

Above and below the disk, we define a density and pressure stratification that is in hydrostatic equilibrium with 
the gravity of the primary, a so-called {"}corona{"},
\begin{equation}
\rho_{\rm c}=\rho_{\rm a,i}\left(\frac{r_{\rm i}}{R}\right)^{1/(\gamma-1)}\!,\,
P_{\rm c}=\rho_{\rm a,i}\frac{\gamma-1}{\gamma}\frac{GM}{r_{\rm i}}\left(\frac{r_{\rm i}}{R}\right)^{\gamma/(\gamma-1)}\!.
\end{equation}
The parameter $\xi \equiv \rho_{\rm a,i} /\rho_{\rm d,i}$ quantifies the initial density contrast between 
disk and corona. In this paper $\xi = 10^{-4}$.
This initial density distribution is rather stable inside the Roche lobe of the primary.
This is essential, as it allows for initial jet formation that is mainly unaffected by
coronal motion.  
When the jet is launched, the coronal region becomes mass loaded by the outflow and is not affected
anymore by the initial state.
Note however, that a mass accumulation may happen at certain areas in the Roche volume (e.g. 
around the L4 and L5 point).

We prescribe a Keplerian rotation for the matter that crosses the inner boundary.
The rotational velocity profile of the accretion disk is given by
\begin{equation}
  v_\phi (r) = \sqrt{\frac{GM}{r}} 
\begin{cases}
     0,  & {\rm for}\,\, 0 < r < r_0 \\
    \sqrt{1-5\epsilon^2}, & {\rm for} \,\, r_0 < r < r_{\rm i}\\
     \sqrt{1- 2.5\epsilon^2}, &{\rm for} \,\,  r_{\rm i} < r <  r_{\rm out}\\
     0, & {\rm for} \,\,r > r_{\rm out}
\end{cases}
\label{vphi_profile}
\end{equation}
where $r_{\rm i}$ denotes the inner disk radius and $r_0$ the inner radius of the ghost area corresponding 
to the inner boundary condition.

 In our simulations the initial outer disk radius $r_{\rm out}$ is smaller than the size of the
computational domain.
The advantage of this prescription is that due to the weak disk rotation for large radii, no specific treatment is 
required at the outer grid boundary, in particular if the disk radius is smaller than the size of the computational 
domain.
The disadvantage is that the mass reservoir for accretion is limited by the finite disk mass. 
This may constrain the running time of the simulation as soon as the disk has lost a substantial fraction of 
its initial mass \citep{2015ApJ...814..113S}. 

However, since it is essential to treat the accretion process properly, we cannot use a similar strategy for the 
inner boundary and just neglect rotation over there.
We thus make use of the {\em internal boundary} option of PLUTO and  define the boundary values in a way that 
allows to absorb the disk material and its angular momentum and that ensures an axisymmetric rotation pattern in 
the innermost disk area (see \citealt{2015ApJ...814..113S}, appendix).

For the outer boundaries of the computational domain, standard outflow conditions are applied as prescribed by PLUTO.

The orbital period of the binary $T_{\rm b}$ is defined as
\begin{equation}
 T_{\rm b}=2\pi\sqrt{\frac{|D|^3}{G(M_{\rm p}+M_{\rm s})}},
 \label{Orbital_period}
\end{equation}
and the orbital angular velocity as
\begin{equation}
\omega = \sqrt{\frac{G({M_{\rm p} + M_{\rm s})}}{|D|^3}},
\end{equation}
where $D$ is the binary separation.
For example, the orbital period of the binary system for run {\em d150a0} with a binary separation
of $D=150$ is  $T_{\rm b} \simeq 8160$ dynamical time steps.

\subsection{A time dependent gravitational potential}
\label{sec:gravity}
The gravitational potential of a binary system is the well-known Roche potential.
The outflows will be launched deep within the Roche lobe of the primary star.
When the outflow propagates away from the launching area, it becomes more and more affected by the tidal 
forces of the Roche potential, and will finally leave the Roche lobe of the primary.
 
Since the origin of our coordinate system is in the primary, we have to consider the time variation of 
the gravitational potential in that coordinate system.
Below we describe the total gravitational potential that is affecting a mass element in the binary 
system (see Figure \ref{fig:binary_cartoon2}).
Assuming that the position of the secondary at $t=0$ is within the $x$-$z$ plane, its position vector is
\begin{equation}
\vec{D}= \hat{x} D \cos{\omega t} + 
         \hat{y} D \sin{\omega t}\cos{\delta} + 
         \hat{z} D \sin{\omega t}\sin{\delta}.
\end{equation}
The effective potential in a binary system at a point with position vector $\vec{r}( x, y, z )$ is
\begin{equation}
\Phi_{\rm eff} = - \frac{G M_{\rm p}}{|\vec r|} - \frac{G M_{\rm s}}{|\vec{r}-\vec{D}|} 
                 + \frac{G M_{\rm s}}{|\vec{D}|^3}  \left(\vec{r} \cdot \vec{D}\right).
\label{eq:phi_eff}
\end{equation}
The first term in Equation~\ref{eq:phi_eff} is the gravitational potential of the primary,
while the remaining terms describe the tidal perturbations due to the orbiting secondary.
These equations have been used previously by other theoretical studies
(see e.g.~\citealt{Papaloizou1995, 1996MNRAS.282..597L, 2010A&A...511A..77F}).
We notice again that the reference frame for our simulations is {\em not} an inertial frame and that 
the last term in Equation~\ref{eq:phi_eff} - usually denoted as indirect term - accounts for the acceleration 
of the origin of the coordinate system. 

The motion of masses that are moving in the binary system are governed by the total (effective) 
gravitational potential.
From the five Lagrange points that exist, three of them ($L1, L2,L3$) are meta-stable points, while the other 
two ($L4 , L5$) are truly stable points.
In the meta-stable points a small perturbation to a mass distribution element will lead to instability\footnote{We
note that besides the gravity, also the magnetic field and the gas pressure will affect the dynamical stability in 
the $L4$ and $L5$ points.}.
We therefore expect to see specific imprints in the dynamical evolution of the disk and the outflow
when these constituents approach or pass the Lagrange points or the Roche lobe.
Note that in our coordinate system (centered on the primary) the Lagrange points orbit with the orbital 
frequency of the binary.
%

\subsection{Limits of our model approach}
\label{limits}
Here we discuss the limits of our model approach.
First of all, we point out that the main aim of the paper is to investigate the {\em jet launching} from
a circum-stellar disk that is located in a binary system.
Our main aim is not to model a {\em binary system} as such.
The binary system provides the background gravity for jet launching.
Our simulations can be seen as complementary to a large number of studies considering 
hydrodynamical models and simulations of disks in binary stars
(see e.g.~\citealt{1994ApJ...421..651A, 1996MNRAS.282..597L, 1999ApJ...512L.131T} 
for early modelling, and
\citealt{2008A&A...487..671K, 2013MNRAS.434.1946N, 2015A&A...583A.133P, 2017ApJ...838...42B} 
for more recent simulation work).
Without the need of considering the magnetic field and a computational domain that allows for jet 
propagation, these simulations reach much longer physical times, as of now up to 100s of orbital periods
(see e.g.~\citealt{2008A&A...487..671K, 2010A&A...511A..77F, 2012A&A...539A..18M}).

Jet launching implies the necessity of MHD and a large vertical grid extension, while the binarity implies
a fully 3D approach.
As a result, these simulations are numerically expensive, running two months per parameter run
on 256 cores corresponding to $\simeq$ 5000 processors.

With the computational resources and a suitable time frame at hand, we are able to run the simulations about
5000 physical time steps corresponding to about 1000 revolutions of the inner disk.
This is indeed sufficient to follow the jet formation process from inner disk, as the typical jet dynamical
time scale roughly corresponds to the Keplerian time at the launching point.
This enables us to detect any changes in the propagation characteristics of the inner jet easily.
Such a change could indicate a hypothetical jet precession when the jet source moves along its orbit.
The 5000 dynamical time units that defined by the inner disk rotation correspond, however, ''only" to about 
half a binary orbit for the chosen separation and mass ratio.

It is known, however, that tidal effects are expected to have {\em fully} evolved only after some binary orbital
times (see e.g. \citealt{1996MNRAS.282..597L, 2008A&A...487..671K, 2012A&A...539A..18M}).
Thus, in order to be able to see any indication for tidal effects, we have to shrink the binary to a suitable 
size.
Here, we apply a binary separation of 150-200 inner disk radii.

This separation corresponds to about 15-20 AU for young stars, 
or about 2000-3000 Schwarzschild radii for a neutron star disk (see Table~\ref{tbl:2}).
While these values are consistent with some of the observed sources, as for example 
Gliese 86 \citep{2000A&A...354...99Q}, Gamma Cephei \citep{1988ApJ...331..902C},
HD 196885 \citep{2008A&A...479..271C}, or microquasars such as SS433 \citep{2017A&A...602L..11G},
we do not (and we cannot) intend to model specific observed systems.

\begin{table} 
\caption {Physical scaling of examples of potential binary jet sources, based on
our model setup applying a binary separation $D = 150 - 200 r_{\rm i}$.
Here, 
$R*$ is the radius of the star,
$R_{\rm in}$ is the inner disk radius, and
$R_{\rm S}$ is the Schwarzschild radius. 
Naturally these values depend also on the masses of the binary components.}
\begin{center}
\begin{tabular}{cccccccc}
\hline
\hline
\noalign{\smallskip}
 Source            &  $R_{\rm in}$  & $D$  \\
\noalign{\smallskip}
\hline
\noalign{\smallskip}
\noalign{\smallskip}
{ Young stellar objects}    & $ 3R* \simeq $ 0.1 AU           & 15 - 20 AU  \\
{ Cataclysmic variables }   & $ 3R* \simeq 2000 R_{\rm S}$    & $30.000-40.000 R_{\rm S}$\\
{ Neutron star binaries}    & $ 3R* \simeq 15 R_{\rm S}$      & $2.000 -3.000 R_{\rm S}$  \\
     \hline
  \end{tabular}
 \end{center}
\label{tbl:2}
\end{table}

We note that within our model approach the choice of stellar separation and the initial disk radius also puts
a constraint on the available disk mass reservoir.
The disk looses mass by accretion and ejection, and without replenishing the mass that is lost, the disk will
disappear after about one orbital time (in our setup).
We have therefore applied a disk as large as possible, meaning being constraint only by the Roche sphere.
This might be in slight contradiction with classical papers 
\citep{1977MNRAS.181..441P, 1977ApJ...216..822P, 1994ApJ...421..651A},
however, the difference in size is only a factor of two.
Later studies although applying a different mass ratio also apply larger disks 
\citep{2007MNRAS.376...89T,2005MNRAS.359..521P, 2012A&A...539A..18M}. 

\section{Results and discussion}
We now present the results of simulations runs that apply a different binary separation,
a different inclination between the initial accretion disk mid-plane and the binary orbital 
plane, or a different binary mass ratio (see Table \ref{tbl:1}).

The major advance to our previous paper \citep{2015ApJ...814..113S} is that we now consider a 3D time-variable 
gravitational potential.
We thus take into account the tidal effects of the secondary ''orbiting" the jet source.
All simulation runs have been performed for a considerable fraction of the binary orbital period, 
the latter corresponding to more than 5000 dynamical time steps, depending on mass ratio and binary separation.

For comparison we have run also a simulation in the hydrodynamic limit that serves as a test case for the 
initial disk structure.
The disk remains stable over several 100s rotational periods until 3D tidal effects disturb the initial structure.
As expected, outflows could not be form by this setup.

\begin{figure*}
  \centering
\includegraphics[width=18cm]{\figurepath/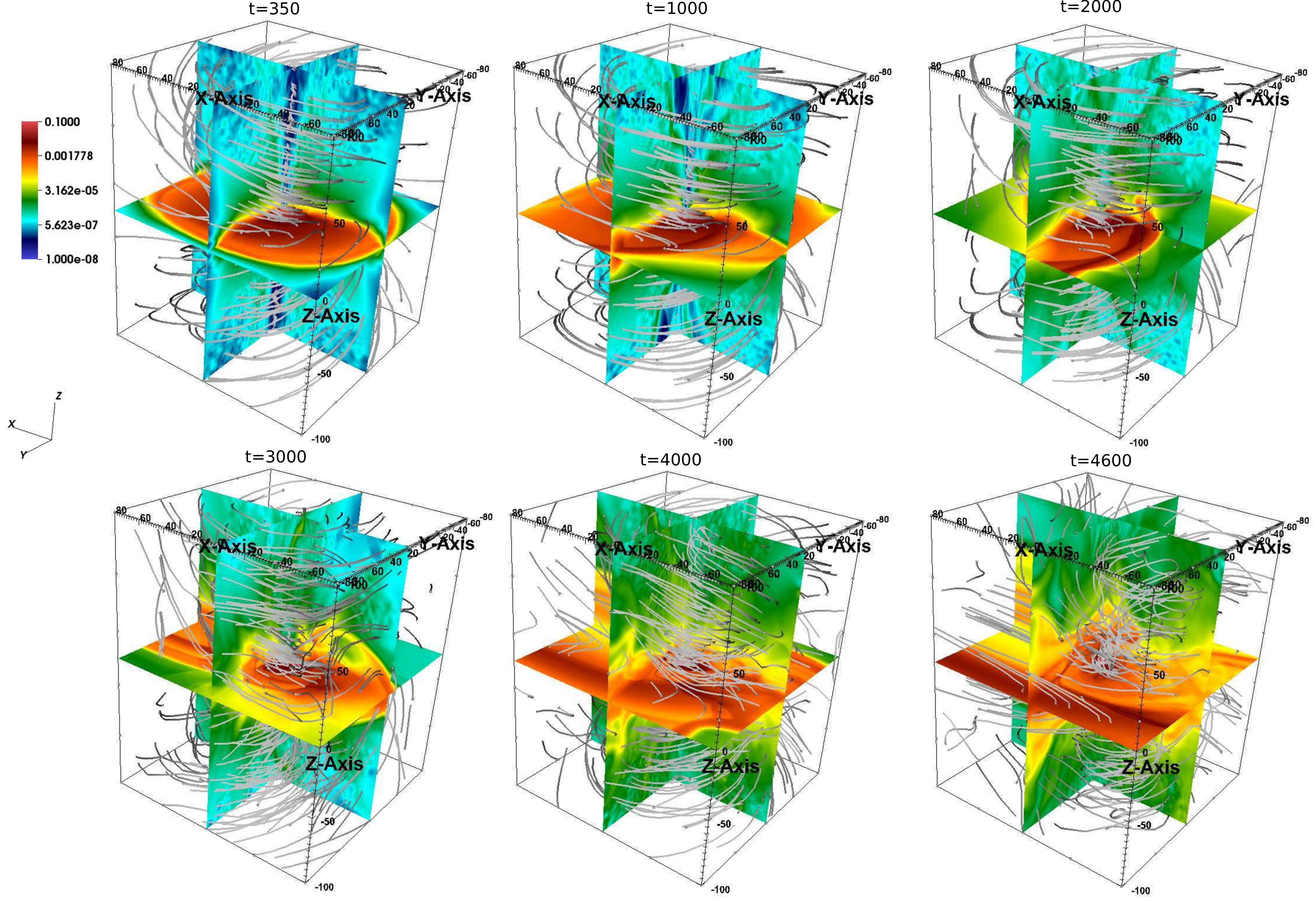}
\caption{ Three-dimensional evolution of simulation run {\em d150a0} with binary separation $ D=150 r_{\rm i}$ 
     and no inclination 
     between the disk and the binary orbital plane.  
     Shown are slices of the mass density (color bar) at times $t = 530, 1000, 2000, 3000, 4000, 4600$ 
     and magnetic field lines (grey).}
\label{fig:ref_3slices}
\end{figure*}

\subsection{Co-planar orbital planes}
\label{COPLANAR}
We first discuss simulation run {\em d150a0} with an orbital plane co-planar to the initial disk plane.
We consider this run as reference run and will compare it with the simulations applying an inclined 
binary orbit and also to the 2D simulations we have published earlier.
This run lasted for about 5000 dynamical times.

\begin{figure*}
  \centering
\includegraphics[width=18cm]{\figurepath/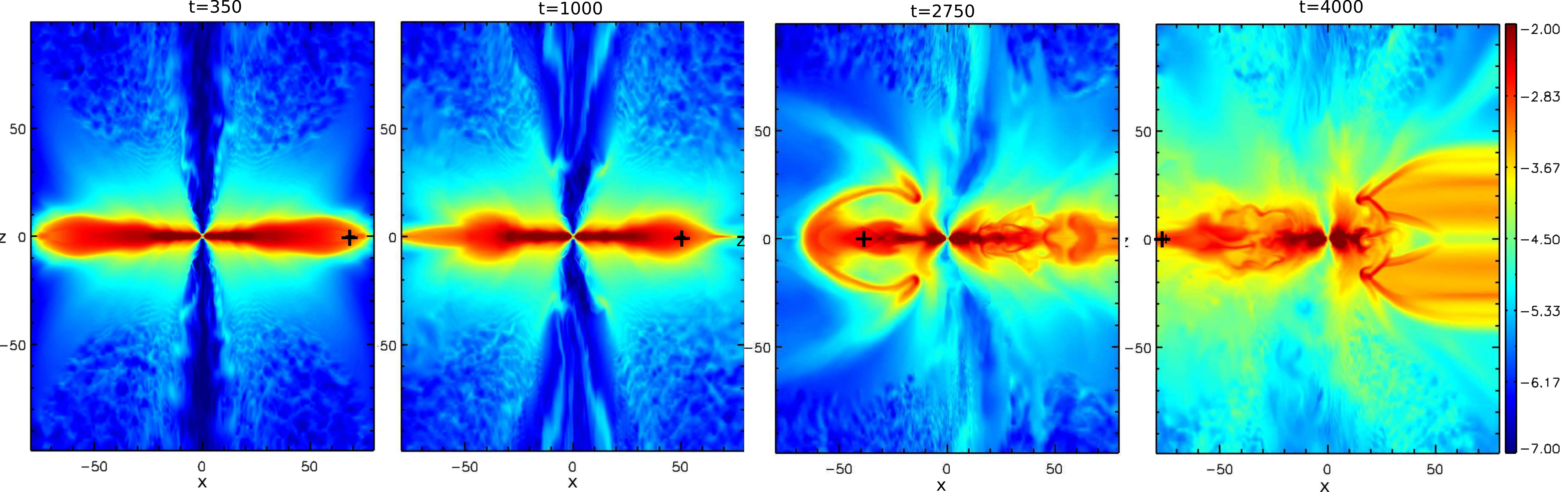}
\caption{
Reference simulation run {\em d150a0} with binary separation $ D=150 r_{\rm i}$ and no inclination between the 
disk and the binary orbital plane.
Shown are two dimensional slices of the mass density in the $x$-$z$ plane for $t= 530, 1000, 2750, 4000$. 
The ``$+$''symbol  indicates to the location of the inner Lagrange point L1 of the binary system projected 
into $x$-$z$ plane.
}
\label{fig:dq150a0_slices1}
  \centering
\includegraphics[width=18cm]{\figurepath/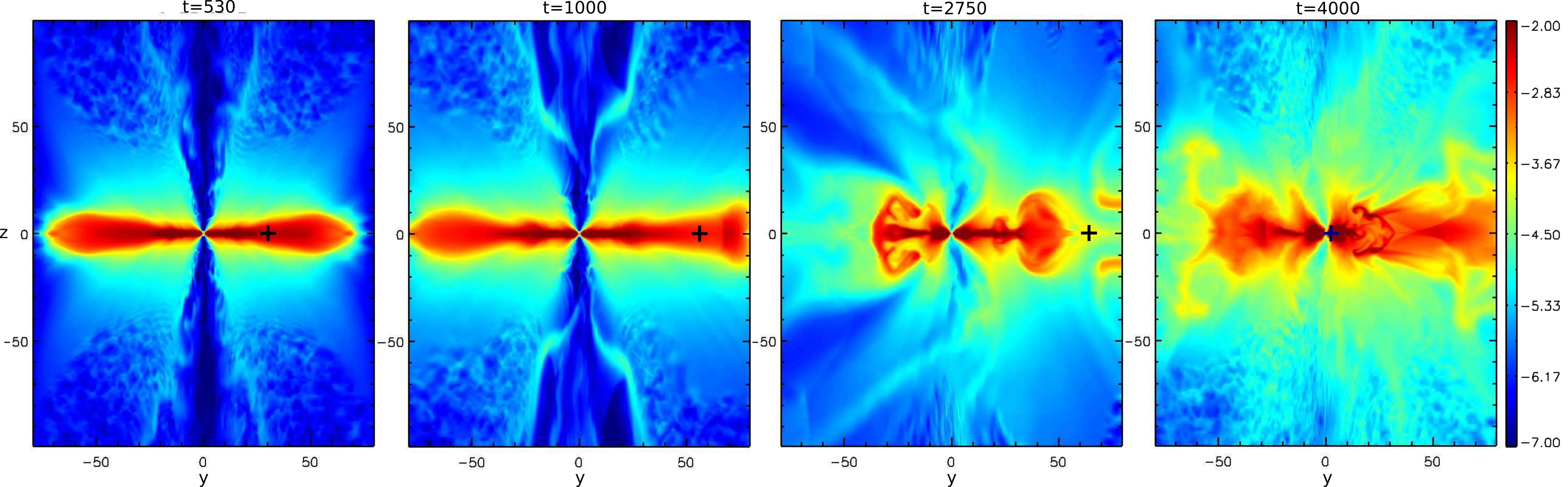}
 \caption{ 
 Same as Figure~\ref{fig:dq150a0_slices1}, but for the $y$-$z$ plane.
 }
\label{fig:dq150a0_slices2}
  \centering
\includegraphics[width=18cm]{\figurepath/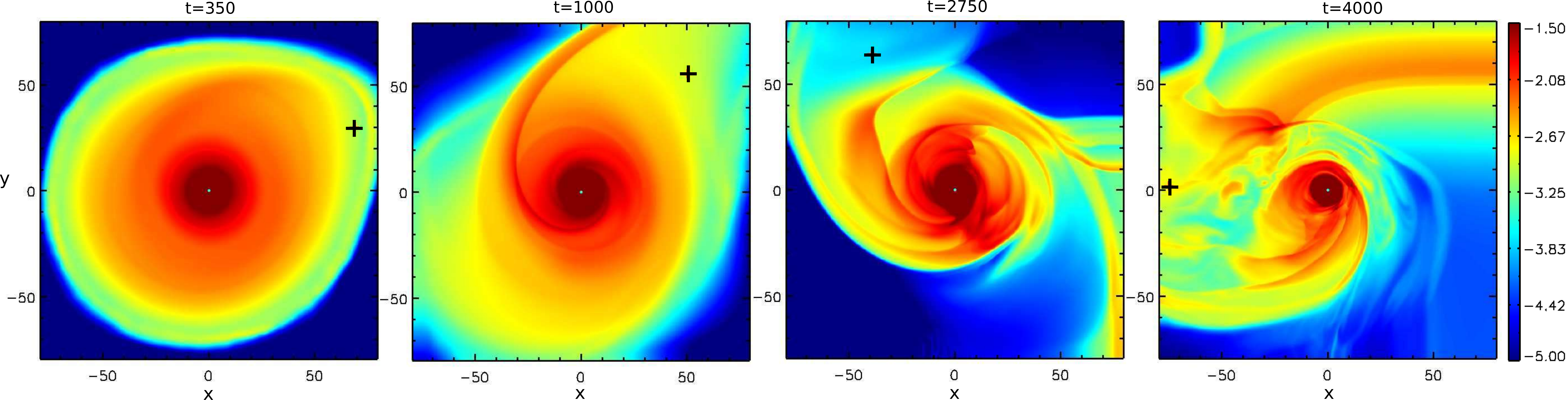}
 \caption{
 Same as Figures~\ref{fig:dq150a0_slices1} and \ref{fig:dq150a0_slices2}, but for the $x$-$y$ plane
 (initial equatorial plane of the accretion disk).
 }
\label{fig:dq150a0_slices3}
\end{figure*}

The global evolution is shown in Figure \ref{fig:ref_3slices} where we plot three-dimensional slices of the 
mass density for times $t= 530, 1000, 2000, 3000. 4000, 4600$, over-plotted with field lines.
We clearly see that a well structured and continuous outflow is launched from the disk.
The disk, the jet and the magnetic field evolve rapidly and change from a smooth structure to a rather 
disturbed pattern.
The details of the evolution is seen better in 2D slices of the simulation.
Figures \ref{fig:dq150a0_slices1}, \ref{fig:dq150a0_slices2}  and \ref{fig:dq150a0_slices3}
show slices of the mass density 
in the $x$-$z$ plane, the $y$-$z$ plane and the $x$-$y$ plane, respectively.
The ''$+$" symbol indicates the (projected) position of the inner Lagrange point $L1$ at each time.

We find that the initial pattern of both the disk and the jet is symmetric to a high degree, 
again demonstrating the quality of the initial setup and of the inner boundary conditions.
However, over long time, i.e. a substantial fraction of the orbit time, the symmetry of the system 
weakens substantially.

Essentially, the upper and lower hemisphere evolve symmetrically with respect to the equatorial 
plane.
As the secondary moves in the disk mid-plane, it induces an asymmetry only in horizontal direction
(it moves from right to left around the vertical axis of the coordinate system in 
Figures~\ref{fig:dq150a0_slices1} and~\ref{fig:dq150a0_slices2}).
Since the binary orbital plane and the accretion disk are co-planar, the material in the upper and 
lower hemisphere feels the same tidal forces.
Overall, this leads to a hemispherically symmetric evolution of the system.

In contrary, the ''left" and ''right" hemispheres feel different tidal forces, which leads to an 
asymmetric evolution in these hemispheres.
This can clearly be seen by comparing the left and right part of the computational domain
for the snapshots taken at different time (in either $x$-$z$ or $y$-$z$ slices).
Essentially, the asymmetry is visible for both the disk structure and the outflow.

The slice in the disk mid-plane shows further interesting features (Figure~\ref{fig:dq150a0_slices3}).
Prominently visible are the ''spiral arms" that start forming at time $t=500$ and essentially grow 
from outside to the inside.
At time $t=2000$ a clear two-arm structure of the outer disk has developed, and a third arm
is just forming.
This point in time corresponds to about 320 inner disk revolutions, corresponding to a 
quarter of the binary orbital period.
Around $t=2000$, the spiral arms extend down to about $r\simeq30$.
 
At this radius the Keplerian period is only about $1000\,t_{\rm i}$, so disk material 
moving at this radius with Keplerian speed have performed two revolutions\footnote{Note that 
at large radii the orbital velocity around the primary is a consequence of the binary potential 
and will thus deviate from a Keplerian orbit}.
We note that the sense of rotation is the same for the disk material, the spiral arm 
pattern, and also for the secondary\footnote{The motion of the secondary and its position angle can be 
recognized from the position of the ''$+$" symbol in the figures that indicates the position of $L1$.}.

Figure 5 shows how the initial disk becomes truncated.
The disk radius at $t=1000$ is about half the radius of L1.
This is consistent with the literature
\citep{1994ApJ...421..651A, 2007MNRAS.376...89T,  2008A&A...487..671K, 2012A&A...539A..18M, 2013A&A...554A..43D,
2015A&A...583A.133P, 2015MNRAS.452.2396M, 2017ApJ...841...29J}.
In all these works the tidal truncation happens inside the Roche lobe.
Depending on the disk physics applied (these papers were mostly dealing with hydrodynamics and were considering 
viscosity) the truncation radius is typically 0.4-0.5 the binary semi-major axis
\citep{1977ApJ...216..822P, 1994ApJ...421..651A, 2000NewA....5...53B}.
We note that the mass located outside the Roche lobe is lost for accretion to the primary,
and also for ejection as a jet.

The dynamical evolution we have discussed for the disk structure is partly seen also in the outflow.
In particular, we mention the jet material that is passing the Roche lobe in {\em vertical} direction.
Clearly, this material is more and more affected by the gravity of the secondary.
As a consequence, it changes its path of motion away from the initial direction of ejection\footnote{We note that 
all figures in this paper are drawn according to the orbiting coordinate system of the primary.
The true motion, projected on the sky would be measured in the center of mass coordinate system.}.

Since for this simulation the binary orbit is co-planar with the accretion disk, we do not see any changes 
in the disk alignment.
As the jet is ejected from the inner disk also, we also do not see a change in the jet rotational axis.
However, as discussed just above the jet dynamics evolves towards a left-right asymmetry, being 
affected by the tidal forces of the binary.

\subsection{A highly inclined orbit}
We now consider run {\em d150a30} with a binary separation is $D = 150$
and an initial inclination between accretion disk and binary orbit of $30 \degr$.
This run serves as an extreme case with both a large inclination angle and a small separation.
We therefore expect a rapid evolution of non-axisymmetric effects for the disk and the jet.

We have performed this simulation for 4500 dynamical time steps corresponding to more than half of 
the binary orbital period.
As mentioned above, the simulation run time is limited by the mass reservoir of the disk, as 
the disk mass is depleted by accretion and ejection.
Due to the small disk radius (limited by the L1), also the initial disk mass is small, 
and does not allow for a much longer simulation time (see our discussion in  Section \ref{limits}).

Although it would be possible in principle to compensate the mass loss by mass injection\footnote{For example,
a physically constrained mass source could be located in the L1.}, 
such a procedure may break the initially axisymmetric evolution by construction.
Therefore, we do not apply such a procedure in the current study.
Nevertheless, run {\em d150a30} does last long enough in order to develop a number of asymmetric features
that evolve in the disk/jet system.

\begin{figure*}
  \centering
\includegraphics[width=18cm]{\figurepath/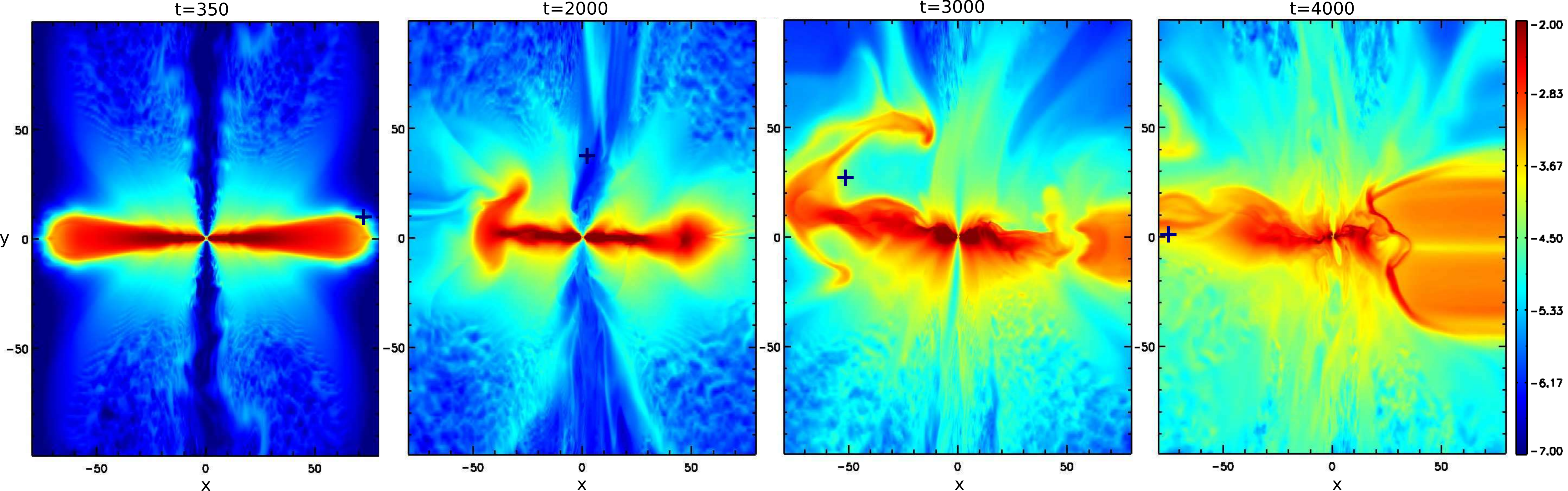}
\caption{
Simulation run {\em d150a30} with binary separation $ D=150 r_{\rm i}$ and $30 \deg$ inclination between the 
disk and the binary orbital plane.
Shown are two dimensional slices of the mass density in the $x$-$z$ plane for $t= 350, 2000, 3000, 4000$. 
The ``$+$''symbol  indicates to the location of the inner Lagrange point L1 of the binary system projected 
into $x$-$z$ plane.
}
\label{fig:dq150a30_slices1}
\includegraphics[width=18cm]{\figurepath/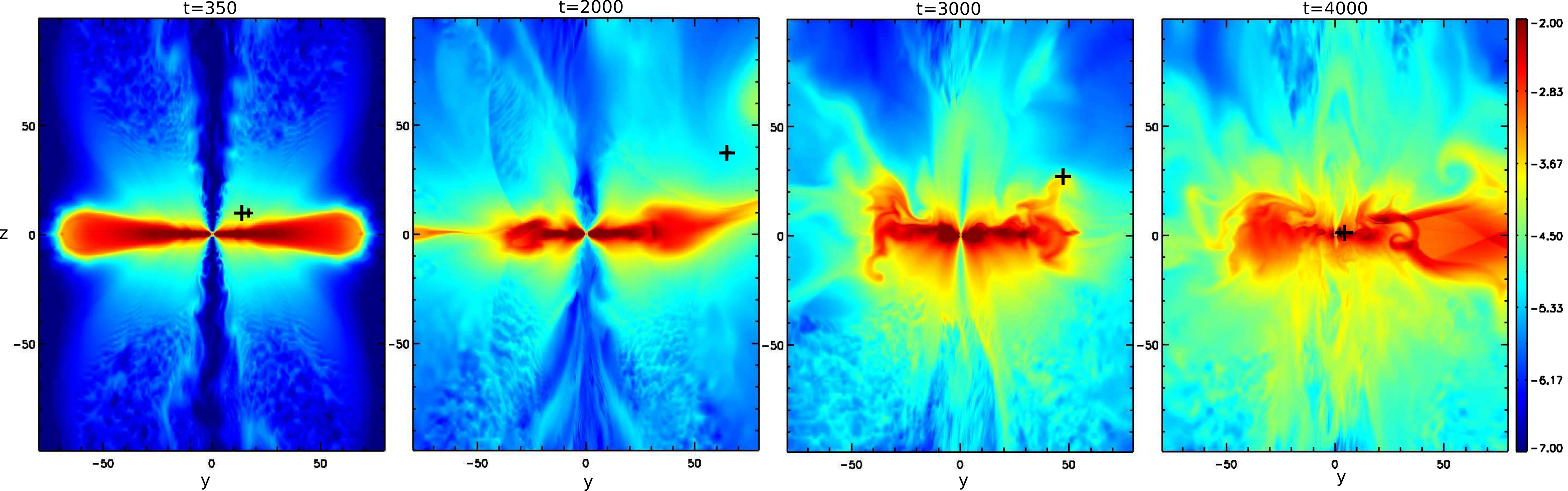}
 \caption{
 Same as Figure~\ref{fig:dq150a30_slices1}, but for the $y$-$z$ plane.
 }
\label{fig:dq150a30_slices2}
\includegraphics[width=18cm]{\figurepath/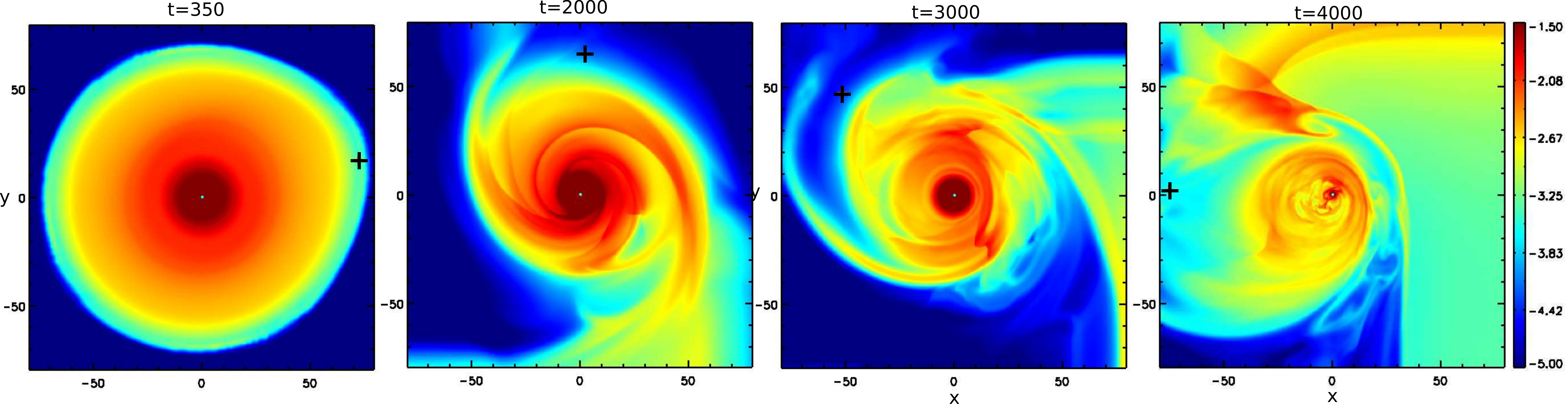}
\caption{
Same as for Figures~\ref{fig:dq150a30_slices1} and \ref{fig:dq150a30_slices2} but for the $x$-$y$ plane 
at $z=0$ (initial accretion disk mid-plane) and for times $t= 350, 1040, 2000, 3000, 4000, 4500$.
Note that at times $t=1000, 3000$ the L3 point is located at distance $r\simeq 75$ from the origin just inside 
the lower left (right) corner.
}
\label{fig:d150a30_rhoslices3}
\end{figure*}
  
In Figures~\ref{fig:dq150a30_slices1} and \ref{fig:dq150a30_slices2} we show slices of mass density 
in the $x$-$z$ plane and in the $x$-$y$ plane for simulation run {\em d150a30} at different time.
We find that after about 500 dynamical time steps a global, 3D non-axisymmetric pattern evolves from 
the symmetric initial setup.
The disk-jet system evolves rapidly, so that after about $t=2500$ the disk structure has completely 
changed.

We first discuss the disk evolution, focusing on the disk area only (see Figure~\ref{fig:d150a30_rhoslices3}).
We note that the initial disk has a axisymmetric structure and is also symmetric in both hemispheres.

\begin{figure*}
\centering
\includegraphics[width=18cm]{\figurepath/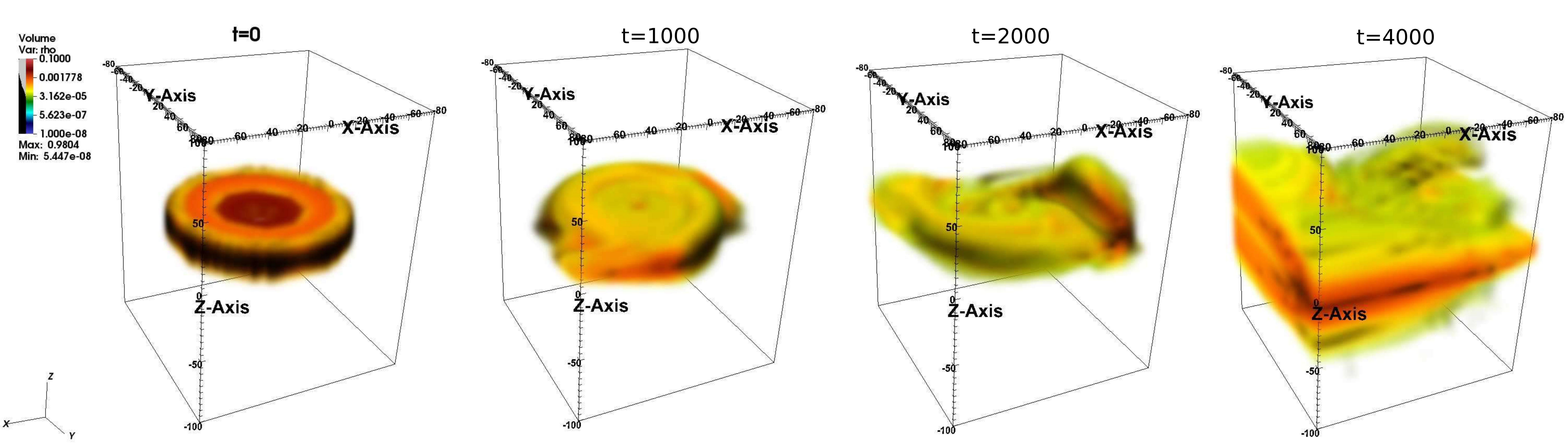}
\caption{
Disk evolution for simulation run {\em d150a30} considering a highly inclined orbital plane. 
Shown is the 3D volume rendering of the mass density (threshold $\rho > 10^{-4}$).}
\label{fig:d150a30volume_rho}
\end{figure*}

When we start the simulation, the secondary is located at position $(x=150, y=0, z=0)$
and reaches the position $(x=0, y=129, z=75)$ after a quarter of an orbit.
Thus, the L1 is initially located at $(x=75, y=0, z=0)$.
After time $t\simeq 500$ the symmetry of the system is slightly broken.
The is visible e.g. in Figure~\ref{fig:d150a30volume_rho} where we show a volume rendering of the disk mass 
density in the $x$-$y$ plane.
Overall we see the formation of spiral arms forming in the disk and a disk structure that is not smooth 
anymore.

In difference to our previous work we see the spiral pattern rotating with the sense of the disk rotation 
and the orbital motion.
In the previous paper with a fixed gravitational potential and considering shorter simulation times, 
a spiral pattern was initiated, but it was not rotating.
The present result is consistent with the orbital motion of the secondary and reflects the effect of 
the time-varying effective gravitational potential implemented in the model.

After $t \simeq 2500$ we see the outer part of the disk beginning to inflate. 
The inflated part of the disk is growing in size and is clearly seen at later evolutionary stages $t > 4000$
(Figure~\ref{fig:d150a30volume_rho}).
We believe that this inflation arises from the fact that the outer disk is located close to the Roche lobe.
As a consequence, it is strongly affected by tidal forces that will destroy the initially Keplerian
disk structure.
Also the vertical gravitational forces are weaker in comparison to a Keplerian disk, so the gas
pressure initially prescribed for the disk will be able to expand the disk easily.

This evolution is similar to that we show in the previous section for the simulation with co-planar orbits.
However, here, the disk stability is even more disturbed by tidal forces due to the higher inclination of 
orbital plane.
Thus, apart from these effects that arise from the local force-balance, also the global dynamical disk structure 
will be perturbed giving rise to warps in the disk.

Our results concerning the circum-stellar disk evolution are in agreement with previous hydrodynamical simulations
of binary systems, also considering a misalignment between the disk and the binary orbit 
(see e.g.~\citealt{1996MNRAS.282..597L, 2010A&A...511A..77F, 2013MNRAS.434.1946N, 2015A&A...583A.133P}).
These works investigate the disk stability and the conditions under which the inclined disk may undergo a rigid 
precession or does form disk warps.
\citet{1996MNRAS.282..597L} find warps and a rigid precession in an inclined thin disk within a binary system and,
in particular, that a very thin disk may be severely disrupted by differential precession, and therefore
cannot survive. 
Similarly, \citet{2010A&A...511A..77F} study how the detailed disk structure that arises in a misaligned binary 
system depends on disk parameters as viscosity or disk thickness and investigate the conditions that 
lead to an inclined disk that may be disrupted by strong differential precession.
 In particular, they find that for a disk thickness $h = 0.05 $
the warp that forms effectively allows the disk to precess as a rigid body with little twisting. 
These disks seem to align with the binary orbital plane on a viscous evolution time.
Thinner disks of higher viscosity develop significant twists before achieving rigid-body precession.

In our simulations we observe the clear impact of the tidal forces on the disk evolution
that lead to asymmetric features like warps and even in some cases lead to a disk disruption - in agreement
with the literature, but now investigated for jet launching MHD disks.

In Figures \ref{fig:dq150a30_slices1}  and  \ref{fig:d150a30_rhoslices3}
we observe that at later evolutionary stages a flow of low density material establishes outside the Roche lobe
(but close to the disk plane),
and then moves inwards where it is obviously disturbing the disk structure and the existing spiral arm pattern.
We believe that this flow is caused by the combination of effects.

Firstly, the spiral arms are growing and thus affecting a larger area inside the disk structure
probably because the inner disk becomes lighter and lighter with time as loosing mass due to 
accretion and ejection. 
Secondly, the disk is inflating in the outer part (in particular close to the L1) and this part of the
disk mass will be loaded into the area outside the Roche lobe.
The further dynamics of this flow is then determined by the effective potential of both stars, eventually 
resulting in the flow structure we just observe.
It is, unfortunately, impossible to predict analytically the dynamics of such a flow.
But as the code considers the full MHD dynamics of the flow we are confident that the flow structure we observe
is in fact a physical effect.

In summary, we think that the combination of warp formation inside the disk in connection with the orbital 
motion of the secondary (disk inflation) results in forming a flow of matter out of the outer part of the disk
that is filling the space around the disk.
Jet launching is initially not affected, but when the launching source - the disk - becomes weaker with time,
also jet formation is suppressed.

\begin{figure*}
\centering
\includegraphics[width=18cm]{\figurepath/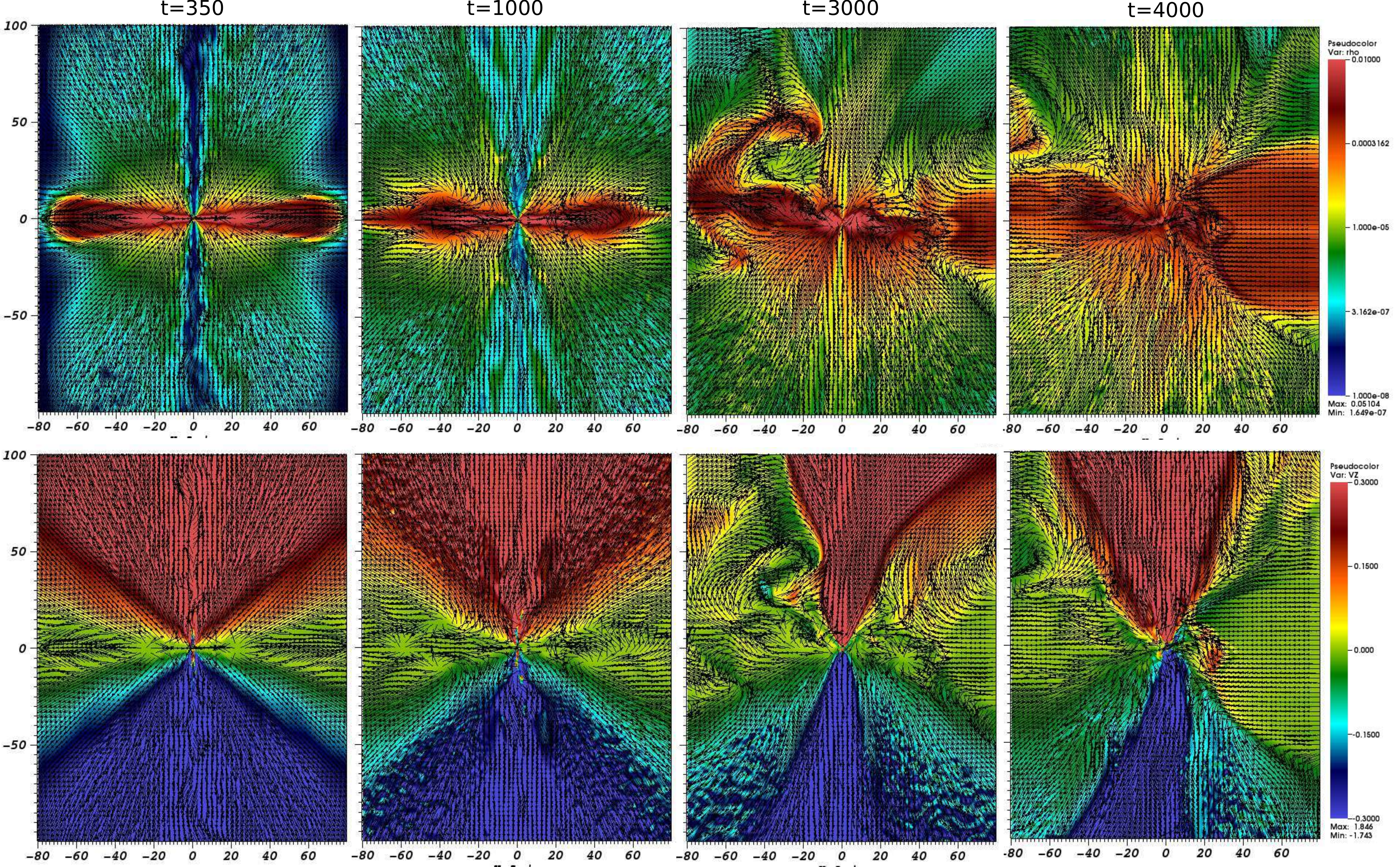}
\caption{Poloidal velocity maps. Shown are two dimensional slices of the vertical velocity $|v_z|$ (top) 
and the mass density (bottom) in the $x$-$z$ plane for  run {\em d150a30} at time  $t= 350, 2000, 3000, 4000$. 
The arrows indicate the (normalized) velocity vector field of the $\vec{v}_p $ projected in $x$-$z$ plane.}
\label{fig:dq150a30_vzvectore}
\end{figure*}
 
Like the disk, also the jet undergoes a rapid time evolution.
Figure \ref{fig:dq150a30_vzvectore} shows 2D slices of the vertical velocity $v_z$ (top) and the mass density (bottom) in the $x$-$z$ 
plane for simulation {\em d150a30} at time $t= 350, 2000, 3000, 4000$. 
In addition the arrows are indicating to the velocity vector field projected in $x$-$z$ plane. 
 Note that the blue central spine in the upper-left panel is {\em not} the jet. 
The jet-wind is launched from all over the disk surface is constitutes of the ''yellow-green"
material (representing low velocities) close to the disk, that is ejected and accelerated to large 
distance from the disk surface.

We find that a strong and axisymmetric bipolar jet is formed during the early evolution.
For $t> 1500$ we see that the jet propagation starts to deviate from its initial straight motion
that is perpendicular to the initial disk plane.
The direction of jet propagation is influenced by two effects.
Firstly, the accretion disk, thus the jet source, aligns towards the binary orbital plane.
Therefore, as a consequence, the direction into which the jet is launched changes as well. 
We find that the deviation from the alignment with the initial rotational axis increases with time.
Secondly, the jet direction of propagation is affected by the global tidal forces, thus the orbital motion 
of the secondary.
These tidal forces of the binary system will influence the jet propagation more and more when it 
is propagating further away from the launching point,
in particular, when the jet leaves the Roche lobe of the primary.

Essentially, we find that at late evolutionary stages both the accretion and the ejection pattern is 
disrupted and no outflows should be expected from such kind of launching geometry
with a large inclination between disk and orbital plane.

\subsection{Impact of the (initial) inclination angle}
We now consider additional simulation runs and compare the disk and jet evolution for different initial 
inclination angle.

Binary orbits that are inclined against the disk mid-plane has been studied previously.
For example \citet{1983MNRAS.202.1181P} and \citet{1997LNP...487..182P} have studied a non-planar disk in close 
binary system analytically, whereas \citet{2010A&A...511A..77F} applied hydrodynamical simulations in order to 
study the detailed structure and evolution of disks in misaligned binary systems.
However, jet launching or disk winds were not considered in this work.

\begin{figure*}
  \centering
\includegraphics[width=18cm]{\figurepath/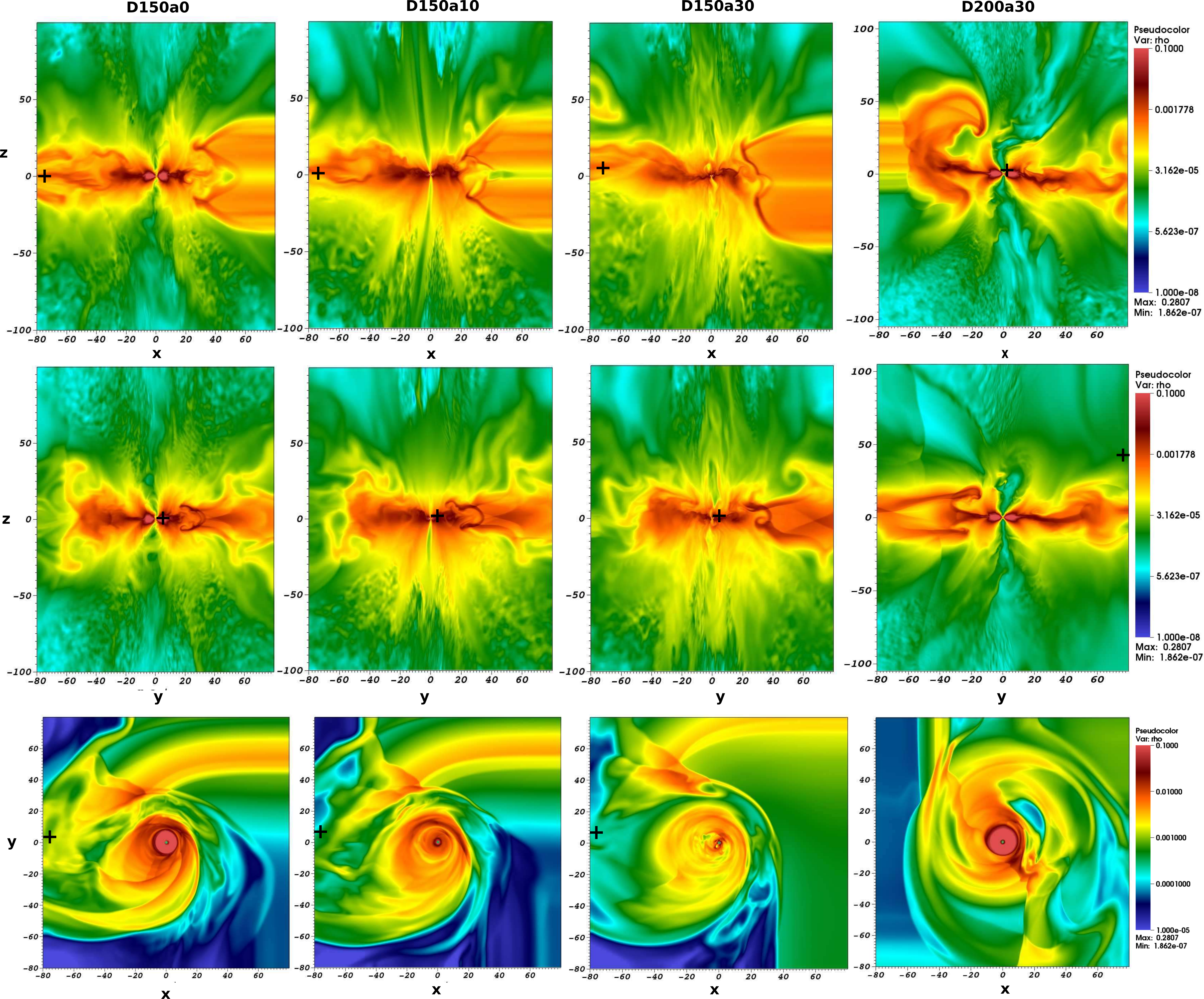}
\caption{
Comparison of the four simulation runs {\em d150a0}, {\em d150a10}, {\em d150a30}, and {\em d200a30} (from left to right)
applying different initial inclination between the disk plane and the orbital plane and a different binary separation. 
Shown are 2D slices of the mass density in the $x$-$y$ , $x$-$z$ and $y$-$z$ plane (from top to bottom)
around time $t=4000$.
See Table~\ref{tbl:1} for the simulation parameters.}
\label{fig:rho_copm_4000}
\end{figure*}

We first consider the mass density distribution in different runs at a specific, late evolution time.
Figure~\ref{fig:rho_copm_4000} shows 2D slices of the mass density distribution
in the $x$-$z$ , $y$-$z$ and  $x$-$y$ plane, around $t= 4000$ for different runs 
applying a mass ratio of unity (see Table~\ref{tbl:1}).
The global evolution in all four runs clearly follows a similar pattern.
However, we observe that in the cases considering a large initial inclination angle (the last two snapshots) the 
alignment of the disk changes strongly.
In contrary, in run {\em d150a10} with an initial inclination angle of $10\degr$, the initial alignment of the  
disk mid-plane is only slightly changed over time.
Note that while tidal forces on the disk are similar for different disk inclinations, the {\em tidal torque} is
different, thus explaining our findings for the differences in disk alignment.

We further see that the deviation of the jet axis from its initial direction is larger in case of a larger 
initial inclination angle.
Moreover, the internal structure of both the disk and the jet changes more over time.
In particular, for run {\em d150a30} the outflow launching mechanism is clearly stalled
at late evolutionary stages, beyond which we do not detect any well structured bipolar jets.
Essentially, we conclude that there exists a critical angle between the accretion disk and the binary 
orbit beyond which the launching of typical jet is suppressed by 3D tidal forces.
For the simulation setup investigated here, the critical angle is between 20 and 30\degr.

It is interesting to note that all jet sources found so far carry a strong magnetic field and are 
surrounded by an accretion disks.
However, no jets have been found for Cataclysmic Variables and are extremely rare for pulsars
although both kind of systems may also have a strong magnetic field and also host an accretion disk
(see for example \citealt{1996Natur.382..789M} or \citealt{1994ApJ...424..955S} for disk accretion,
or \citealt{2017RAA....17...10W} for the detection of a magnetic field).

Similar to cataclysmic variables, jet from neutron stars have not been observed in general.
There are few exceptions such as the Vela pulsar \citep{2013ApJ...763...72D} and, as mentioned above, SS\,433. 
Also the Crab pulsar shows a time-variable, elongated feature that could be interpreted as a jet \citep{2013MNRAS.436.1102M}.

In summary, (almost) no jets have been observed from these numerous and extremely well observed binaries
\citep{1998MNRAS.297.1079K, 2004A&A...422.1039S}.
An explanation for this observational fact has been suggested by \citet{Fendt1998} who consider a certain degree 
of {\em axisymmetry} as essential ingredient for jet launching.
Our present simulations support this idea.

 \begin{figure*}
 \centering
  \includegraphics[width=18.5cm]{\figurepath/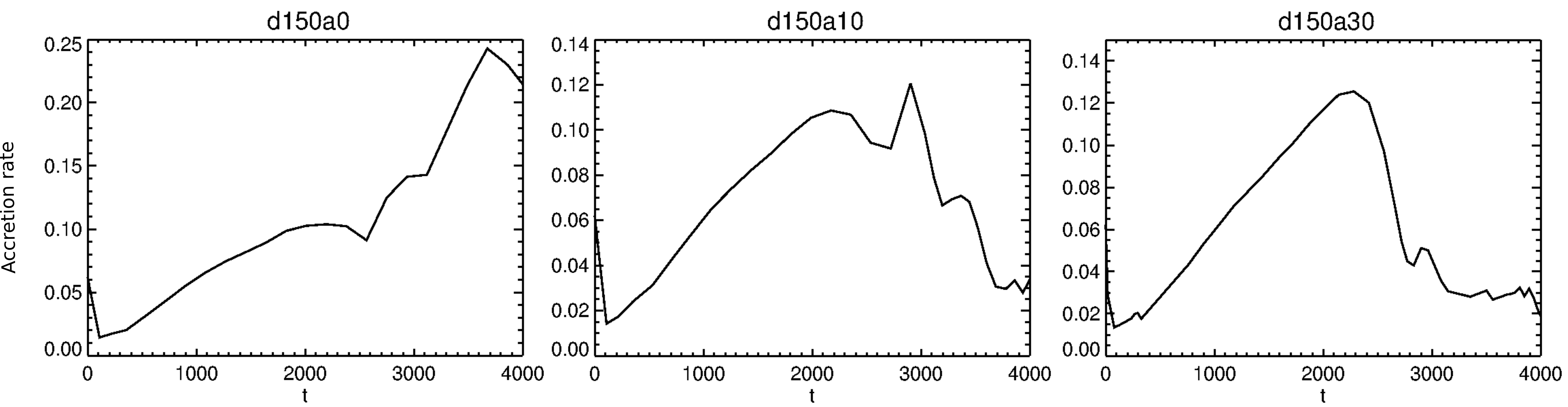}
 \caption{
 Time evolution of the accretion mass flux.
 Shown are the mass fluxes in the disk integrated along $r=2$ and for three (initial) thermal scale heights $h$ for 
 the simulation runs {\em d150a0} (left),  {\em d150a10} (middle)and  {\em d150a30} (right).
 }
\label{fig:acc_comp}
 \includegraphics[width=18.5cm]{\figurepath/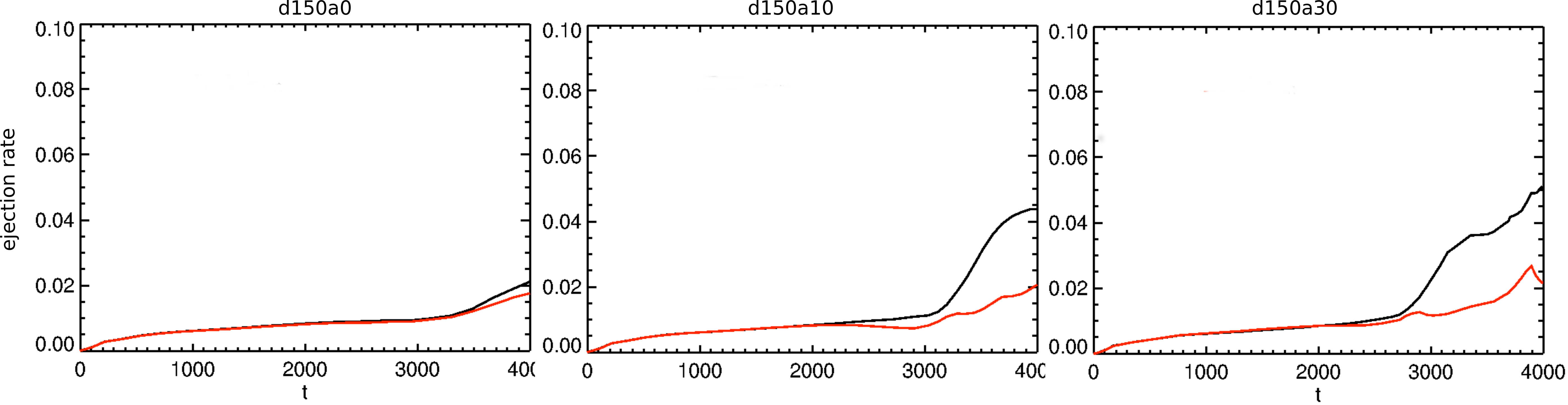}
 \caption{
 Time evolution of the bipolar outflow mass flux. Shown are the mass fluxes $\dot{M}$ for the jet (upper hemisphere; in red) 
 and the counter jet (lower hemisphere, in black ) for the simulation runs {\em d150a0} (left) {\em d150a10} (middle) and  
 {\em d150a30} (right). 
The number values were integrated far from the disk at altitude $z=90$ .
}
\label{fig:ejec_comp} 
\end{figure*}

A good way to quantify the launching efficiency and also the system asymmetry
is to evaluate the mass fluxes in the disk and in the outflows.
Figure~\ref{fig:acc_comp} shows the the evolution of the disk accretion flux, integrated at disk 
radius $r = 2$ over three (initial) thermal scale heights $H$ for the simulation 
runs {\em d150a0}, {\em d150a10} and  {\em d150a30}.
Although in all cases the global evolution of the system looks similar, the simulation applying a 
co-planar orbit with the disk has on average a higher accretion rate.
In all cases, the time evolution of the accretion flux peaks at later time steps. 
We understand this peak as due to the onset of 3D disk instabilities, the spiral wave, or warps that enhance the 
angular momentum transport and thus facilitate accretion.

Figure~\ref{fig:ejec_comp} shows the evolution of the ejected mass flux for jet and counter jet for 
the simulation runs {\em d150a0} (top) {\em d150a10} (middle) and  {\em d150a30} (bottom).
We have integrated the asymptotic value for the outflow mass fluxes far from the disk at $z=\pm 90$.
As a result, we observe that the vertical mass flux in both hemispheres is similar and only later
the hemispheric symmetry is broken.
It is strongly indicated that the asymmetry between the mass flux integrated in each hemisphere is 
higher for the simulation runs {\em d150a10} and {\em d150a30}, thus runs those with an asymmetric initial
setup.
This is understandable as the hemispheric symmetry of the gravitational potential is broken by the
secondary that is located initially in the upper hemisphere.
We expect that when we would have been able to run our simulations over much longer time, e.g. several orbital periods,
the average outflow mass flux in both hemisphere would be on average comparable.
%
\subsection{Impact of the binary separation}
We now consider additional simulation runs and compare the disk and jet evolution for a different 
binary separation.

In simulation run d200a30 - considering a larger binary separation-the L1 is located at 100$r_{\rm i}$ 
(outside the domain) and the orbital period of the binary is substantially longer, $T_{\rm b} = 12,560 t_{\rm i}$.
This simulation was run for about 5000 dynamical time steps corresponding to about one third of the binary orbit.

In Figure 11 (last row) we show $x$-$y$ slices of the mass density distribution of  run {\em d200a30} that we 
can use to compare the disk structure and with its spiral arms to the other simulations.
Clearly, the initial axisymmetric disk structure is disturbed by tidal forces as in the other runs. 
However, as these forces are weaker due to the larger separation, the disk looks more regular and ''roundish{”}. 
Compared to simulation {\em d150a30} the inner disk is less dissolved (if we compare both disks for $ r < 10$).
The ''roundish{”} part of the disk is also larger for the present simulation than for the simulations applying a smaller 
binary separation. 
This is as expected as L1 is now outside the computational grid.

Besides the weaker tidal forces, also the time scale after which we may identify tidal effects has increased as the 
orbital period is longer.
We find a factor $\simeq 2$ for the increase of this time scale.
Overall, the impact of tidal effects on jet launching critically depend on the binary separation\footnote{Applying a 
larger separation, we would not be able to disentangle these tidal effects for jet launching in reasonable CPU time}.

In Figure \ref{fig:rho_copm_4000} we observe a ``warp'' forming in the disk (see the snapshots of mass density 
in the  $x$-$z$ and $y$-$z$ plane).
These warps are only forming in simulations that start off with an initial inclination between the disk 
mid-plane and the binary orbit.
The warps are stronger in simulation run {\em d150a30} that applies the largest initial inclination angle
and a small separation, thus reflecting a large tidal influence by the secondary.

\subsection{Impact of the binary mass ratio}

In order to investigate the impact of the binary mass ratio on the disk-jet dynamical evolution
we have also run simulation with a mass ratio $M_{\rm s}/M_{\rm p}$ of 2 and 0.5, denoted as {\em d150m2.0}
and {\em d150m0.5}, respectively.
The mass ratio obviously affects the size of the Roche lobe of the primary, thus the tidal forces 
on disk and jet, and the (orbital) time scale of the evolution (see Table~\ref{tbl:1}).

In our setup, the orbital period of the binary increases, $T_{\rm b} = 6664, 8162, 9424$, for decreasing 
mass ratio $M_{\rm s}/M_{\rm p} = 2.0, 1.0, 0.5$, respectively.
Therefore, to investigate the evolution of a low mass ratio system requires more CPU time.
On the other hand, the tidal effects by a high mass secondary on a low mass jet source is expected
to be higher, and thus to appear earlier with respect of the orbital period.

%

\begin{figure*}
  \centering
\includegraphics[width=15cm]{\figurepath/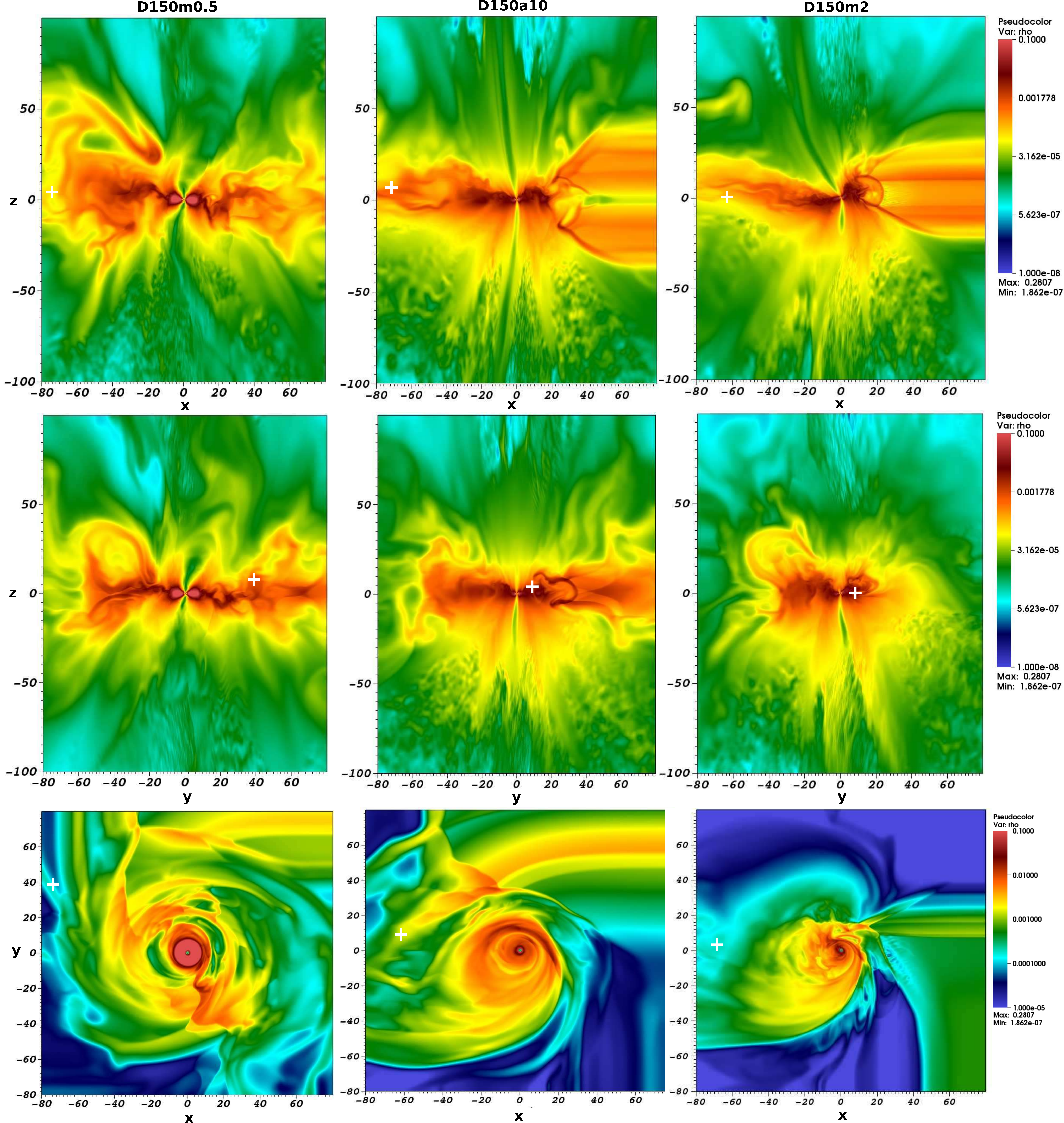}
\caption{
Comparison of launching simulations with different binary mass ratio. 
Shown are 2D slices of the mass density in the $x$-$y$, $x$-$z$ and $y$-$z$ plane (from top to bottom)
around time $t=4000$ for the three simulation runs {\em d150m0.5}, {\em d150a10}, {\em d150m2} 
(from left to right).
The ``$+$''symbol indicates the projected location of the inner Lagrange point L1
(the L1 is located inside the domain for simulation {\em d150m2}). 
}
\label{fig:rho_q_4000}
\end{figure*}

 \begin{figure*}
 \centering
  \includegraphics[width=18.5cm]{\figurepath/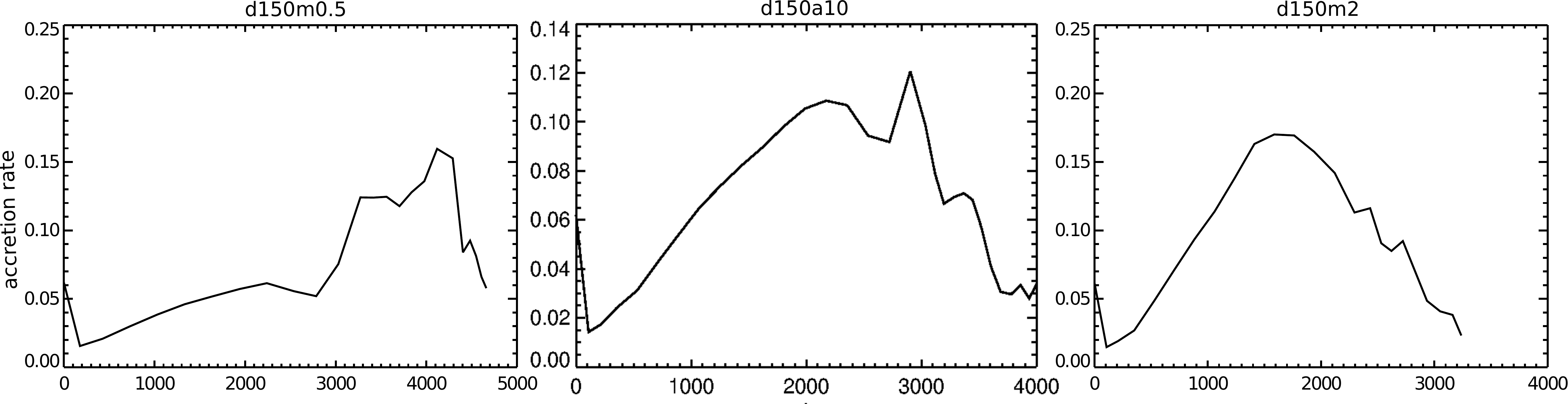}
 \caption{
 Time evolution of the accretion mass flux.
 Shown are the mass fluxes in the disk integrated along $r=2$ and for three (initial) thermal scale 
 heights $h$ for the simulation runs {\em d150m0.5} (left), {\em d150m2} (middle), and {\em d150a10} 
 (right), applying a different binary mass ratio.
 }
 
\label{fig:acc_comp_q}
 \includegraphics[width=18.5cm]{\figurepath/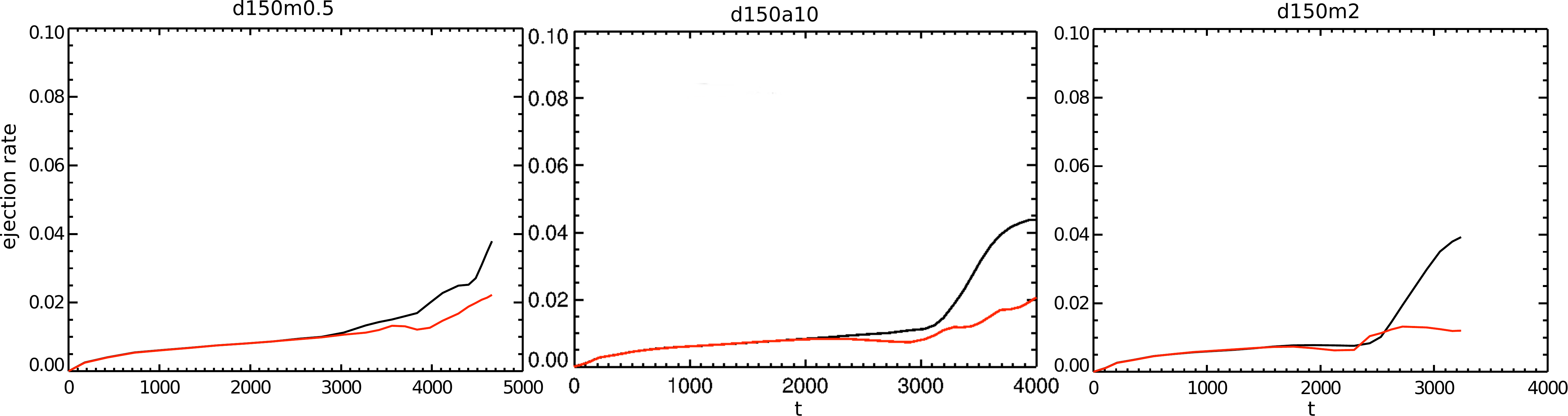}
 \caption{
 Time evolution of the bipolar outflow mass flux. 
 Shown are the mass fluxes $\dot{M}$ for the jet (upper hemisphere; in red) and the counter jet (lower 
 hemisphere; in black ) for the simulation runs {\em d150m0.5} (left), {\em d150a10} (middle), and  
 {\em d150m2} (right), applying a different binary mass ratio. 
The mass fluxes are integrated far from the disk at altitude $|z|=90$.
}
\label{fig:ejec_comp_q}
\end{figure*}

In Figure \ref{fig:rho_q_4000} we compare 2D slices of the mass density in the $x$-$y$, the $x$-$z$, and 
the $y$-$z$-plane (from top to bottom) at $t \simeq 4000$ for the three simulation runs 
{\em d150m0.5}, {\em d150a10}, {\em d150m2} (from left to right) applying a different mass ratio.
We find that the disk structure in run {\em d150a0.5} is less disturbed in comparison with the disk in run
{\em d150m2} for which the disk has deviated from a round structure substantially and is also truncated 
at a smaller radius.
This is clearly an effect caused by the larger tidal forces in run {\em d150m2} with a large mass 
ratio $M_{\rm s}/M_{\rm p} = 2.0$.

We find as well that the jet is more deflected from its original direction along the axis for the higher
mass ratio (see in particular the $x$-$z$-slice for{\em d150m2}), and also that the deflection 
is stronger in the upper hemisphere in which the secondary is located.

We have also investigated the impact of the mass ratio on the global evolution of the system
by considering the mass fluxes in the disk-jet structure. 
This is illustrated in Figure~\ref{fig:acc_comp_q} showing the mass fluxes in the disk integrated 
at $r=2 $ and for three (initial) thermal scale heights $h$ for the simulation runs
{\em d150a10} (left), {\em d150m2} (middle) and {\em d150m0.5} (right) applying different mass ratio.

Essentially, we find a delay in the accretion evolution of run {\em d150m0.5} compared to other two 
runs.
Also, the accretion rate in simulation {\em d150m2} is slightly higher than for {\em d150a10}.
This is probably due to the stronger and earlier formed spiral arms in the disk and thus a more 
efficient (tidal) angular momentum removal from the disk.

As for the accretion rate, in Figure~\ref{fig:ejec_comp_q} we compare the evolution of the ejected mass 
flux for jet (in red) and counter jet (in black).
Here we find that the run {\em d150m0.5} shows a weaker asymmetry in the outflow mass flux in the upper 
the and lower hemisphere compared to the other simulations.
The explanation is similar as for the accretion rates. 
In simulation {\em d150m0.5} the tidal forces of the secondary are weaker and cannot disturb the flow
symmetry as much as for the other simulations.

\subsection{Jet / disk precession?}
 The original motivation for these simulations was the question whether we can find indication 
for a disk or jet precession, similar to what the observations suggest (see Sect.~2).

Above we have discussed our findings that whenever a misalignment is present between the 
binary orbital plane and the (initial) disk plane, we observe a re-alignment of the disk and,
as a consequence, subsequently also a re-alignment of the outflow axis.
The question remains whether the re-alignment we observe will actually evolve into a constant 
precession of the disk and jet axis.
 Precession would imply an orbital motion of the jet axis around the initial jet axis (the z-axis)
along a cone with a certain opening angle.

\begin{figure*}
\includegraphics[width=16.5cm]{\figurepath/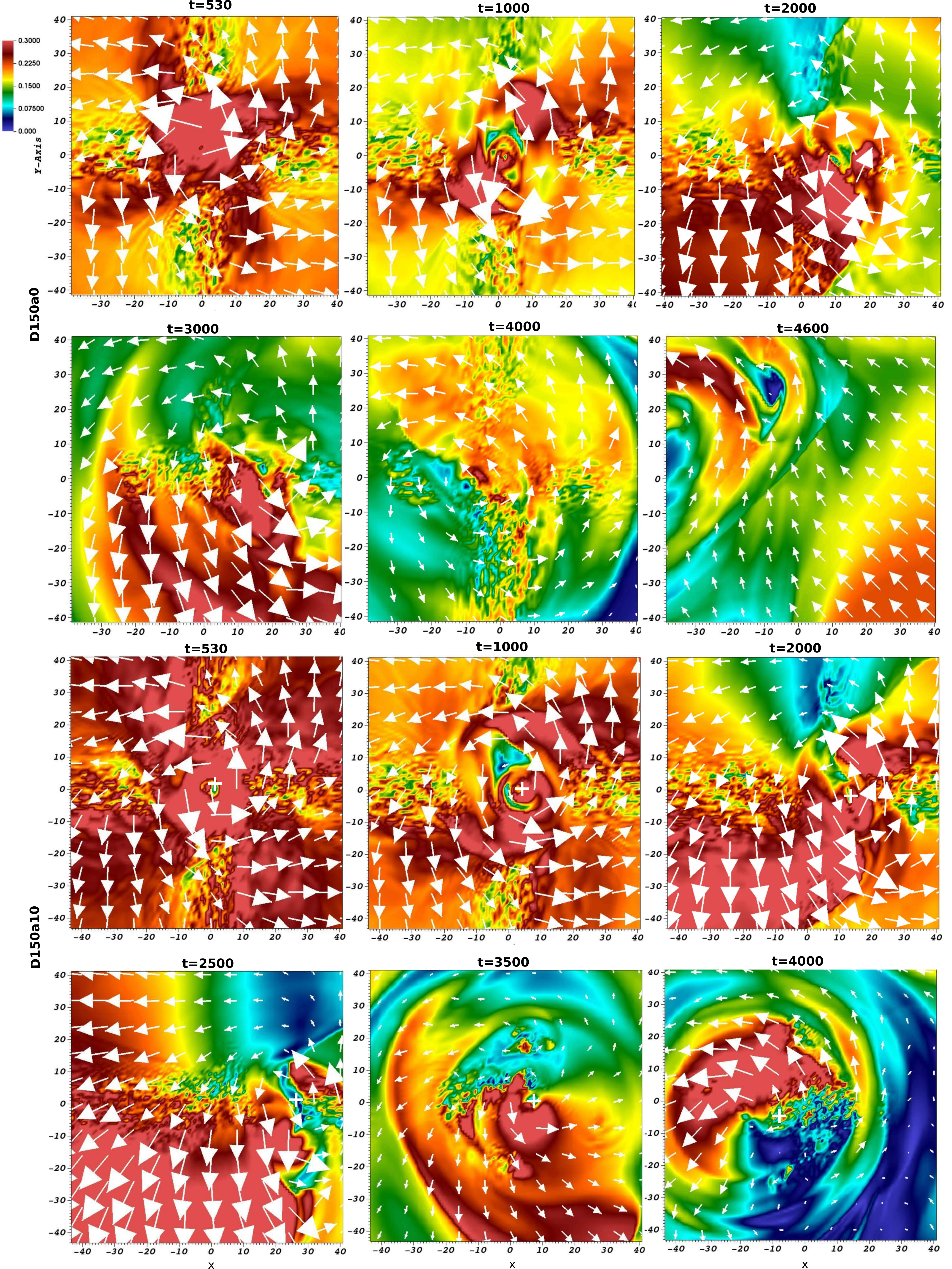}
\caption{
Rotational evolution in the outflow.
Shown are slices of the magnitude of the $x$-$y$ velocity, thus $\sqrt{v_x^2+v_y^2}$ (color) in the $x$-$y$ plane at $z=60$ 
for the two simulation runs {\em d150a0} (upper two rows) and  {\em d150a10} (lower two rows) at different times.
While this velocity pattern is dominated by the toroidal velocity in the outflow (rotation),
a small contribution of the radial velocity of the outflow (projected in to the $x$-$y$ plane) cannot be 
disentangled (the main poloidal outflow component is vertical).
The arrows indicate the velocity vector field projected into the plane.
The ``+'' symbol illustrates the approximate position of the jet rotation axis.}
\label{fig:allruns_z60_vrot_smooth}
\end{figure*}

Since our simulations consider all gravitational forces present in the system
(the Coriolis force and the time-dependent gravitational potential), we could in principle
find and prove a true precession of the disk and jet.

However, as already mentioned above,
since precession evolves over several orbital time scales, a long run time of the simulation 
would be required.
Typical estimates of the precession time scale of a disk in a binary system due to tidal forces
is of the order of  $T_{\rm P} \simeq 20 T_{\rm b}$ \citep{2000MNRAS.317..773B}.

This is of course a challenge considering a 3D MHD setup and the required numerical resolution.
The typical running time scale of our simulations is about 4000-5000 dynamical time steps, 
corresponding to 20-30 years in case of YSOs or 7-10 years in case of AGN.
This substantially less than the time scale needed for jet precession to evolve fully.
Thus, we would only be able to {\em indicate the onset of precession}.
Nevertheless, after half of an  orbital period of the binary, corresponding to about 500 revolutions of the
jet launching inner disk, 
we expect to see indication of a circular motion of the jet axis that is inclined against the
original jet and disk rotational axis (at $t=0$), and thereby indication of a precession effect.

Figure~\ref{fig:allruns_z60_vrot_smooth} shows two dimensional slices of the {\em projected} velocity 
in the $x$-$y$ plane at $z=60$ for two runs {\em d150a0} and  {\em d150a10}.
We plot the velocity magnitude (color-coded) and the velocity field (arrows).
The panels consider simulations runs {\em d150a0} (rows 1, 2 ), {\em d150a10} (rows 3, 4 ), from 
top to bottom.
Note that what is shown in Figure~\ref{fig:allruns_z60_vrot_smooth} is the projected 
velocity  $\sqrt{v_x^2+v_y^2}=\sqrt{V_r^2+V_{\phi}^2}$ 
that is a superposition of the rotation and the radial flow (for example the radial outflow of the jet).
It is well known that for classical MHD jets that are super-Alfv\'enic, $v_{\rm p} >> v_{\rm A}$, 
we have $v_{\rm p} >> v_{\phi}$, and that far from the source in a collimated jet we expect
$v_z >> v_r$.
Thus, we expect a (projected) radial outflow motion that is, however, not large.
Thus, $v_{\phi} \simeq v_r$, implying that is difficult to disentangle orbital motion from 
radial motion in the jet.

After all, it is not straight forward to determine the exact position of the jet rotation axis.
We estimate the approximate position of the jet axis considering the direction of the velocity
arrows.
In particular, close to the jet axis, the radial outflow speed is expected to be small compared to the 
vertical speed, thus the velocity vectors are related more to the rotational pattern than to the outflow
pattern.
Note, however, that due to the overall 3D gravitational potential and the fact that the the slice at
$z=60$ is close to the Roche lobe, we may also expect a lateral bulk motion of the jet that could have 
a larger amplitude than the expected precession.
If this is true, we would not be able to detect precession.

In Figure~\ref{fig:allruns_z60_vrot_smooth} we have applied the procedures just explained 
and have marked with a ``+'' symbol the approximate position of the jet rotation axis.
Indeed, we clearly observe a feature indicating jet rotation and we also see a clear displacement of the 
jet rotational axis in time clearly in both cases.
In particular, for simulation run {\em d150a10} the jet axis resembles a kind of orbital motion.

After all, 
we may interpret this motion as indication for a jet axis precession and derive a precession angle
for simulation {\em d150a10} around $t= 4000$.
We find a displacement in $x$ and $y$ direction of about $8 r_0$ and $3 r_0$, respectively.
Thus, considering the altitude of the slice taken at $z=60 r_0$ this corresponds to about 
angle $\simeq 8 \degr$ for the displacement.

We use the same strategy for simulations {\em d150m2} and {\em d150m0.5} with similar inclination angle 
of $10 \degr$, but a mass ratio of 2 and 0.5, respectively.
We find that the change in the jet rotation axis position is faster in the run with mass ratio of 2
compared two other runs as tidal effects on the jets are strongest in this case.
In all three runs the offset of the  jet rotation axis relative to its initial position is about a 
few degree.

We emphasize that this interpretation has to be taken with care.
The present simulations last considerably longer than those we published before \citep{2015ApJ...814..113S}.
However, in order to prove a true jet and disk precession, simulation run times of several orbital periods
are needed.
So far we are limited in this respect by the limited disk mass that is available for accretion and ejection, 
and by computational resources.

Finally we notice that at late time steps the velocity fields does not really show a regular rotation pattern
in particular for large radii.
This can be seen well best in runs {\em d150a30} and {\em d200a30} which are set up with a high initial inclination 
angle of $30\,\degr$ between the initial disk and the orbital plane.
We interpret this evolution as given by the high asymmetry in the binary setup and the subsequent 3D tidal
forces which are so strong that they destroy the aligned outflow that has been
established from the disk initially.

\section{ Comparison to observed sources}

In Sect.~3.5. we have discussed the limitations of our model approach.
Mainly due to limited CPU resources we cannot yet run simulations over many
orbital periods, and also cannot provide an extensive comparison of parameter
run.
Keeping this in mind, this section is devoted to a comparison of our simulation
results to observations of jets in binary systems.

As we know from observations, jet precession is indicated in different astrophysical jet 
sources and, in addition, there are jet sources that are indicated as binary systems.
On the other hand, from our simulations we can clearly state that if there is a 
misalignment between the disk orbital plane and the binary orbital plane,
a re-alignment of the disk axis and the jet axis takes place.

In our simulations this misalignment is prescribed with the initial condition.
However, a more realistic approach (but not yet feasible for MHD) would be to
simulate a binary system over long time and {\em evolve} a misalignment ab initio.
This has been done in the literature by applying long-term hydrodynamic simulations
(see e.g.~\citealt{2014MNRAS.440.3532L, 2017MNRAS.469.2834O}). 
In this sense our simulations provide strong and direct indication for jet precession resulting
from dynamics of the accretion disk - as we do launching simulations that physically treat the
transition from accretion into ejection.

A more reliable proof of our findings would follow from long-term simulations over several orbital 
periods, which are currently numerically not feasible.
Unfortunately, the same numerical constraints 
prevent us from actually fitting our simulations to 
observed sources.
The physical orbital period of a young star binary with $R_{\rm in} \simeq 0.1 {\rm AU}$, a separation of 15 $\rm AU$,
and mass ratio of 0.5, 1, 2 is about 34, 42 and 48 years, respectively.
The typical running time of our simulations is 4000-5000 dynamical times, corresponding to
800-1000 inner disks Keplerian times and 20-30 years (see our discussion in Sec.4 of \citealt{2015ApJ...814..113S}).

For comparison, we now discuss the few jet-emitting binary systems for which the essential 
orbital parameters are known.

The first example is the binary micro-quasar 1E 1740.7−2942 for which a semi-major axis about 0.36\,AU and a
mass ratio of 1:5 is estimated \citep{2015A&A...584A.122L}.
Therefore, the size of the Roche lobe has 25\% the size of the semi-major axis.
In \citet{2015A&A...584A.122L} radio maps of 1E 1740.7−2942 were analyzed for five epochs covering the 
years 1992, 1993, 1994, 1997 and 2000, with an angular resolution of a few arcseconds. 
Structural changes in the arc-minute jets of 1E 1740.7−2942 were clearly detected on timescales of roughly a year.
The observed changes in the jet flow suggest a precession period of about 
$490\,$d $\simeq 1.3\,$yrs.
The ratio of precession period $P_{\rm prec}$ to orbital period $P_{\rm orb}$ is thus predicted to 
be in the range of $P_{\rm prec}/P_{\rm orb} \simeq 20-40$. 
With the orbital period of 12.7\,d suggested by \citet{2002ApJ...569..362S} this ratio becomes about 40.
While the mass ratio is not too far from our scaling, the separation of 0.36\,AU would correspond to
$\simeq 10^5$ inner disk radii$R_{\rm in}$ assuming an inner disk radius of 10 neutron star radii.
This is far from our model setup considering a binary separation of $150-200\,R_{\rm in}$.

Another example of jet precession is the X-ray binary SS\,433.
This source is observationally very well constrained.
The orbital parameters and the kinematics of its relativistic jet could be modeled accurately
(see e.g. \citealt{1989ApJ...347..448M, 2002SSRv..102...23C, 2006ApJ...650..338L, 2013ApJ...775...75M, 2013MNRAS.436.2004C} ).
The precession period is 162.5\,d, while
the binary inclination angle is assumed to be $79\degr$, the jet precession angle $20 \degr$,
and the jet nutation angle $6 \degr$ (see \citealt{2013MNRAS.436.2004C}).
%
A mass ratio in the range  of $\simeq 0.3-0.5$ is estimated.
\citet{2010ApJ...722..586S} give a binary separation of $\simeq 60\rsun$ and a Roche lobe radius of $\simeq 28 \rsun$.
While the latter two values are similar to our own geometrical scaling if compared to a protostar, 
the inner disk radius in SS\,433 should be much further in as the central object is a compact star.
Thus, assuming a similar scaling ob $R_{\rm in} \simeq 3-5 R_{\rm star}$ as considered in out setup,
a separation of $60\rsun$ would correspond to a separation of several $10^5$ inner disk radii, way off
our model geometry.

We finally mention the protostellar object Cep\,E ejecting two, almost perpendicular outflows.
This source seems to be a class 1 or class 0 binary and the wiggling structure of one of the outflows
is probably due to precession \citep{1996AJ....112.2086E}.
The latter paper provides a model fit to the binary jet system suggesting a precession angle 
of about $4 \degr$ with a precession period of 400 years,
a mass ratio of about unity, and a binary separation of 68\,AU.
We note that these observationally derived parameters are actually close to our model setup.
However, we hesitate to argue that we can model this source applying our simulations,
although we find a similar precession angle in our simulations with intermediate inclination.
On the other hand, this is consistent with our claim - if the disk-orbital inclination would be 
larger, we would not expect a persistent jet formation.
If the inclination would be smaller, not precession would be expected.
Overall, we may suggest that in Cep\,E the jet launching disk and the orbital plane are inclined 
by some $10-20\degr$.

In summary, our model setup successfully produces different features that are expected from theoretical
studies and are seen in observational data of binary stars, such as disk warps, spiral arms, 
a jet deflection, a bipolar and a horizontal asymmetry of the jet-disk system all indicating
on a jet precession.
Although we cannot fit individual jet sources by numerical constraints in general, our simulations 
have been able to disentangle - for the first time - all tidal effects that affect the jet 
launching process in binary systems.

\section{Conclusions}
We have presented results of fully 3D MHD simulations of jet launching from the circum-stellar disk of a jet 
source orbiting in a binary system.
Extending our previous approach \citep{2015ApJ...814..113S}, the new simulations consider a time-dependent 
Roche potential along the orbit of the disk and the jet.
We consider all tidal forces for the evolution of the jet and the circum-stellar disk
around the primary star.

Our simulations apply the PLUTO code considering Cartesian coordinates.
We run the simulations over a substantial fraction of the binary orbital period, corresponding to 
about 5000 dynamical time steps of the simulation (defined by the rotation of the inner accretion disk).
Modeling the whole binary system including jet launching is beyond numerical feasibility for the 
next future.

We have presented six simulations applying a different {\em binary separation}, a different
{\em inclination} between the initial accretion disk mid-plane and the binary orbital plane,
 and a different binary mass ratio.
In order to be able to measure the expected tidal effects over a reasonable CPU time,
a small binary separation was applied in general.
We have obtained the following results.

(1)
A reference run with an orbital plane co-planar with the initial disk plane was,
resulting in a well structured and continuous outflow launched from the disk.
Both the disk structure and the jet outflow evolve initially highly symmetric, indicating the quality 
of our model setup.
Later-on the axial symmetry of the system is reduced while the bipolar symmetry of the accretion-outflow 
system is still kept. 
This is due to the tidal forces induced by the secondary.
The asymmetry becomes visible only after a substantial fraction of the orbital time scale that 
is much longer than the time scale of outflow.
 
(2) 
A number of 3D features evolve that are common to all our simulations. 
Most prominently, a ''spiral arm" pattern emerges in the disk and grows in time.
The spiral arms start forming rather early at time $t=500$, corresponding to a tenth of an 
orbital period, forming first in the outer disk and then grow from the outside to the inside.
Later, a prominent two-arm structure in the outer disk extends to about 30 inner disk radii.
 The motion of the spiral arm pattern is aligned with the orbital motion of the secondary star.

(3)
Further non-axisymmetric features that we detect are disk warps.
This is known in the literature of hydrodynamic disk simulations in binary systems,
however, it presents a new feature for jet launching simulations and has essential impact for 
the jet stability.
Disk warps form only in simulations that evolve from an initial inclination between the disk plane 
and the binary orbital plane. 
The higher the inclination the larger the disk warp.
Similarly, 
we find that a larger initial inclination results in a stronger re-alignment of the disk plane 
and the jet axis away from their initial position.

(4)
An exemplary simulation run with both a large inclination angle and a close separation shows a
rapid evolution of non-axisymmetric effects for the disk and the jet.
After about 2500 dynamical time steps the disk alignment is changed substantially and, as a 
consequence, the jet propagation direction as well.
The deviation from the initial setup still increases with time and is triggered by tidal effects 
due to the secondary on its orbital path.
We observe that the outer part of the disk starts to inflate, possibly due to its vicinity
to the Roche lobe and thus lacking gravitational support from the primary.

(5)
Simulations with different mass ratio indicate a change of time scales for the tidal forces to 
affect the disk-jet system.
A large mass ratio (a massive secondary) shows a faster evolution of the system and results 
in stronger spiral arm feature, a higher (on average) accretion rate and a more pronounced jet-counter 
jet asymmetry.

(6)
In our simulations we find indication for jet precession.
Deriving the jet axis from the jet rotational velocity pattern, we find for the simulations with moderate 
initial inclination between disk initial plane and orbital plane a displacement
of the jet axis of $\simeq 8\degr$.
This deviation may be interpreted as onset of jet precession.
As precession fully evolves on several orbital time scales only, our findings are so far indications
only.
To follow an established jet precession would require simulations over many orbital times which
is currently without reach for jet launching simulations.

(7)
Comparing for all of our simulation runs the persistence of the jet that is ejected, 
we find that for initial inclination angles larger than $10 \degr$ the jet does not survive the simulation time scale.
A jet is launched and does propagate, but is later destroyed by tidal forces.
This indicates on a critical precession angle beyond which typical jet launching strongly suffers from 
the 3D tidal effects.
This may explain the observational findings that jets are numerous among young stars, where we expect 
the star-disk angular momentum axes to be aligned during the star formation process, 
while jets are rarely seen ejected from compact stars such as pulsars or cataclysmic variables despite 
the presence of an accretion disk and a strong magnetic field.

(8)
We finally mention the limits of the model approach. 
Performing fully 3D, resistive MHD simulations covering a large numerical grid, we are currently limited 
to simulation time scales below one orbital time scale.
In order to be able to disentangle the 3D dynamics in the disk and jet evolution due to tidal effects
of the binary system, we have applied a short binary separation of 150-200 inner disk radii, corresponding
to about 500-1000 stellar radii, depending on the kind of jet source (young star or compact star).
The physics applied follows the standard resistive MHD approach used in jet launching simulation.
Further non-ideal MHD effects or radiation are not yet included, and are not necessary to capture
the main dynamical features of the jet launching.

In summary, our MHD simulations of jet launching from disks in binary systems suggest a critical 
angle between disk plane and orbital plane of somewhat above 10 degrees beyond which a jet cannot 
persistently be formed out of a disk wind.
Our simulations also indicate the onset of jet precession with a precession cone opening angle
of about 8 degrees.
The simulations were performed for stellar separations of 150-200 inner disk radii and are thus
more related to close binary systems.
However, our main findings - the re-alignment of disk and jet axis and the existence of a critical
disk-orbital inclination angle for jet formation can in principle be applied for binary jet sources
in general.

\acknowledgements{
We thank Andrea Mignone and the PLUTO team for the possibility to use their code.
We thank Rony Keppens and Bhargav Vaidya for valuable comments.
We acknowledge helpful comments by two unknown referees that have contributed to a better 
presentation of our results.
This work was partly financed by the SFB 881 of the German science foundation DFG,
by the Max Planck Institute for Astronomy,
and by the Institute for Research in Fundamental Sciences (IPM).
Our simulations were performed on the Theo cluster of the Max Planck Institute for Astronomy
and the Hydra and Cobra clusters of the Max Planck Society.
}

\appendix

\section{Non-Axisymmetric jet formation in 3D - BP mechanism?}
\label{BP_binary}

The critical inclination angle for a jet launching magnetic field line of $60\degr$ is a famous
criterion derived by \citet{1982MNRAS.199..883B}.
It has been modified in the presence of thermal pressure forces \citep{1992ApJ...394..117P}.
Here we discuss a further modification in case a binary gravitational potential.

The effective gravitational potential of a point mass and a Keplerian motion of the footpoints $r_0$
of the magnetic field lines is
\begin{equation}
\begin{split}
  \Phi(r,z)= -\frac{GM}{r_0} \left[0.5\left(\frac{r}{r_0}\right)^2 
                                   + \frac{r_0}{\sqrt{r^2+z^2}}\right] 
\end{split}
\end{equation}
\citep{1982MNRAS.199..883B}.

\begin{figure}
  \centering
\includegraphics[width=6cm]{\figurepath/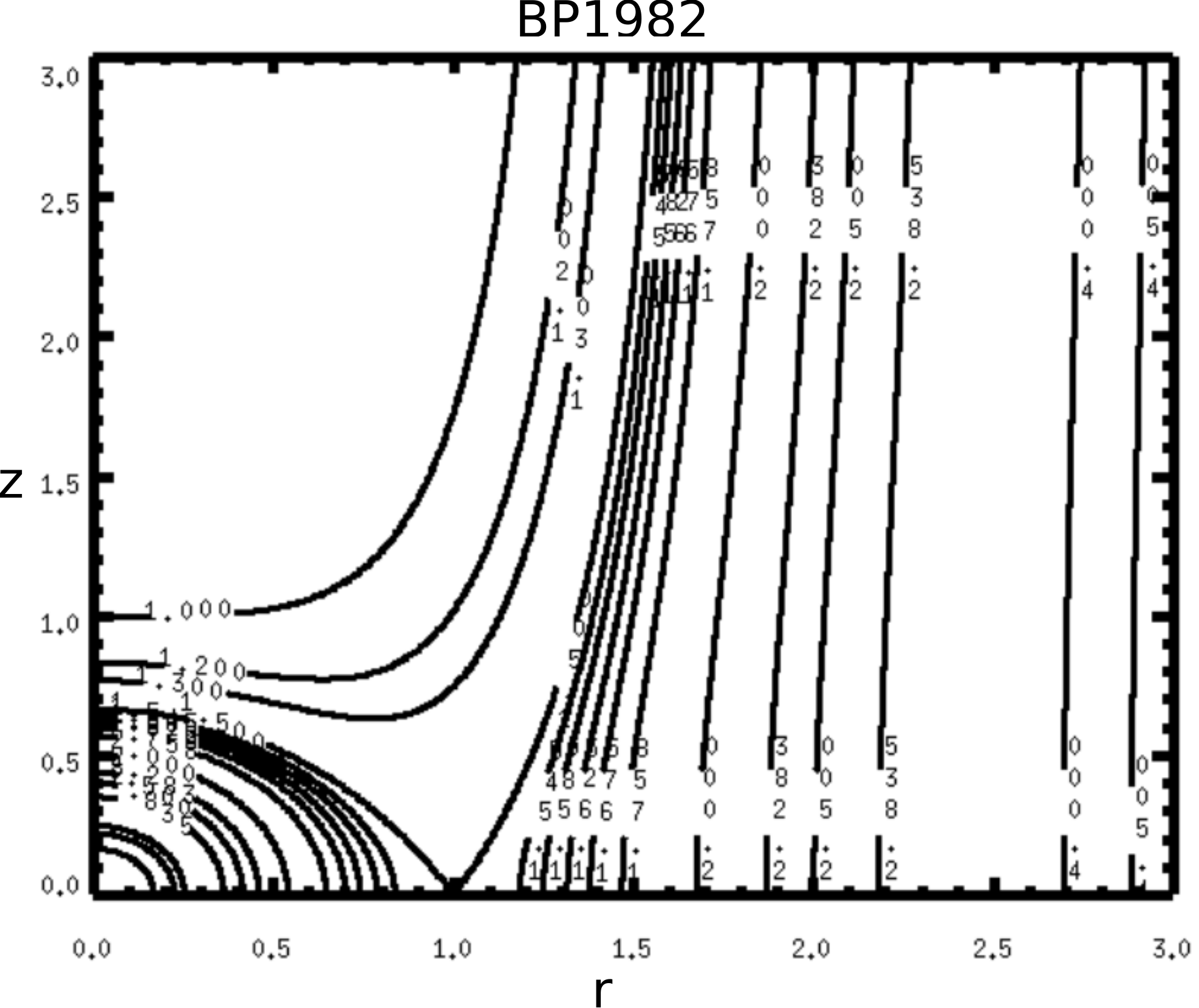}
\caption{
Effective gravitational potential for magnetic field lines corotating with the Keplerian angular velocity (anchored
in an accretion disk). Here, $r_0 =1$.
Shown are arbitrary equipotential surfaces, $\Phi(r,z) = {\rm constant}$, chosen such that a cusp is seen at $r=r_0$
\citep{1982MNRAS.199..883B}.
}
\label{fig:cnbp1982}
\end{figure}

For comparison this is shown in Figure~\ref{fig:cnbp1982}.
Material on field lines emerging from $r_0$ in outward direction and inclined by less than $60~\degr$ towards the equatorial plane,
is unstable against outward magneto-centrifugal acceleration. 
Material on field lines emerging from $r_0$ in inward direction and inclined by less than $60~\degr$ towards 
the equatorial plane,
is unstable against infall.

In case of a binary system, the equipotential surfaces for a mass element co-rotating with the 
magnetic field line rooted in the Keplerian disk  at $r_0$ are represented by the effective 
binary potential, thus by the Roche potential.
In the coordinate system originating in the primary (at a specific time and specific position) this is
\begin{equation}
\begin{split}
  \Phi(r,z)=-\frac{GM}{r_0}\left[0.5\left(\frac{r}{r_0}\right)^2 + \frac{r_0}{\sqrt{r^2+z^2}}+\frac{r_0}{\sqrt{(r-D)^2+z^2}}
  + \frac{r_0}{D^3} \left(r-\frac{D}{2}\right)^2\right].
\end{split}
\end{equation}
Note that in the original treatment by \citet{1982MNRAS.199..883B} self-similarity allows to 
apply Figure~\ref{fig:cnbp1982} to any choice of $r_0$.
In our case, we have a fixed length scale in the system that is given by the binary 
separation.
Therefore, the critical angle for magneto-centrifugal acceleration will change with radius.

\begin{figure*}
  \centering
\includegraphics[width=18cm]{\figurepath/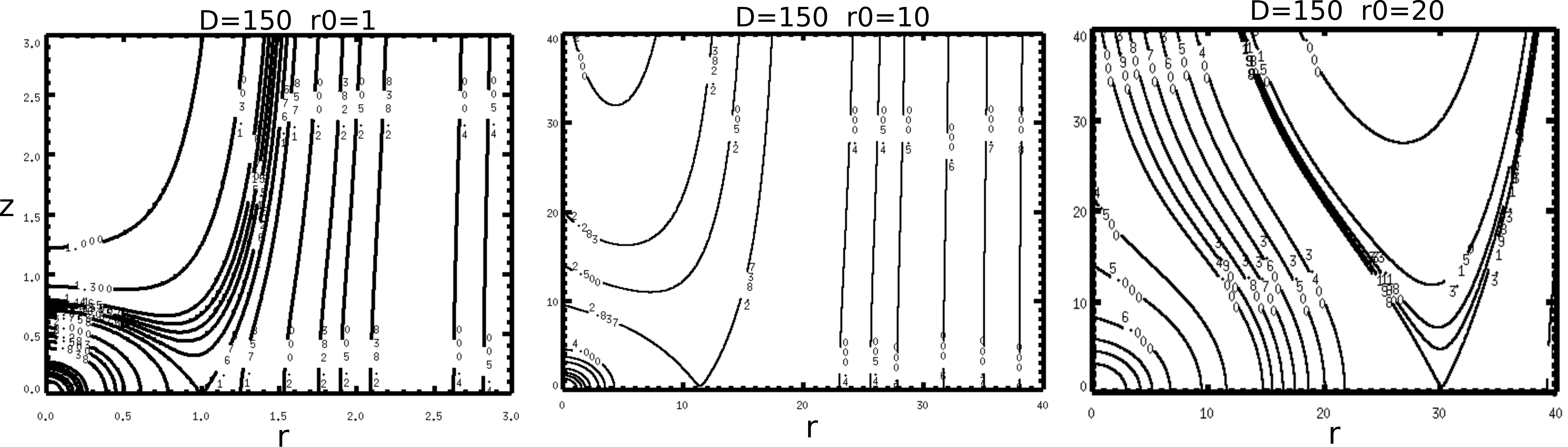}
\caption{Shown are the equipotential surfaces for beads on a wire co-rotating  with the Keplerian 
velocity at $r_0$, and the whole system is part of the binary system. 
Different panel show the contours for different magnetic field footprint $r_0=1,10,20$
and binary separation $D=150$.
}
\label{fig:cnbp}
\end{figure*}

In Figure~\ref{fig:cnbp} we show the equipotential surfaces for ''beads on a wire" co-rotating with
''Keplerian" velocity of the footpoint of the wire, $r_0$.
Note that here with {\em Keplerian} we mean ''in equilibrium with the binary potential".
Different panels represent the equipotential surfaces for the binary separation,
$D=150$ and for different magnetic field line footpoints, $r_0 = 1,10 ,20$.
Considering Fig.\ref{fig:cnbp} we observe that the  equipotential surfaces for the footpoint $r0=1$
are similar to contours  Fig.1 in \citet{1982MNRAS.199..883B} and they show the similar cusp at $r0=1$
and therefore the same critical angle  for the jet launching is .
As we go further out the  equipotential surfaces changes.
Although the contours at $r0=10$ still produce the similar cusp but it is not seen at $r0=10$
and is  shifted to larger radius.
The difference in case with the $r0=20$ is even more , the cusp is shifted to larger radius and also the 
angle  is not $60$ anymore.

Obviously, at the $L1$ point the material is not gravitationally bound to the primary anymore 
and can easily escape vertically (being then bound to the gravitation of both stars.
So, as a first guess, the critical angle for the inner disk is the original $60\degr$ while it
approaches $90\degr$ when approaching the L1 (neglecting the gas pressure).

It is also worth noting an east-west asymmetry in the critical angle. 
As a consequence the Blandford-Payne acceleration will be different depending on the azimutal angle 
around the accretion disk.
So, not only the large-scale outflow propagation will be affected by the Roche potential,
but also the initial acceleration - depending on launching radius and launching position
around the disk.
Only for the innermost part of the outflow, the highly energetic jet, we expect a symmetric
launching and acceleration.

In summary, the initial acceleration mechanism of jets is clearly affected by the 3D potential of the 
binary.
In particular, acceleration along field lines rooted in the outer disk, is somewhat easier.
However, these outflows will not be very energetic.
Due to the lack of vertical gravity the outer part of the disk (close to L1) will be dissolved
more easily.

\bibliographystyle{apj}
\bibliography{mypaperFeb2018}

\newpage

\end{document}